\documentclass[prc,twocolumn,nofootinbib,superscriptaddress,showpacs,aps]{revtex4-1}

\usepackage{amsmath}
\usepackage{amssymb}
\usepackage{cases}
\usepackage{isotope}
\usepackage{siunitx}
\usepackage{rotating}
\usepackage{hyperref}
\usepackage{bm,braket}
\usepackage{multirow,dcolumn,tabularx}
\usepackage[percent]{overpic}

\newcolumntype{.}{D{.}{.}{1.5}}
\newcolumntype{,}{D{.}{.}{0.2}}
\newcommand{\ci}{\text{i}}
\newcommand{\abs}[1]{\left| #1 \right|}
\newcommand{\cg}[2]{
C^{#1}_{#2}}
\newcommand{\sj}[6]{ \begin{Bmatrix}
#1 & #2 & #3 \\
#4 & #5 & #6
\end{Bmatrix}}
\newcommand{\nj}[9]{ \begin{Bmatrix}
#1 & #2 & #3 \\
#4 & #5 & #6 \\
#7 & #8 & #9
\end{Bmatrix}}
\newcommand{\jsc}[1]{\widehat{#1}}
\newcommand{\del}[1]{\delta_{#1}}

\renewcommand{\o}[1]{\bm{\mathrm{#1}}}
\renewcommand{\vec}[1]{%
\ifcat\noexpand#1\relax % check if the argument is a control sequence
\bm{#1}% probably Greek
\else
\textbf{#1}% single character
\fi}

\begin{document}

\title{Shell-model interactions from chiral effective field theory}

\author{L.\ Huth}
\email[Email:~]{lukashuth@theorie.ikp.physik.tu-darmstadt.de}
\affiliation{Institut f\"ur Kernphysik, Technische Universit\"at Darmstadt, 64289 Darmstadt, Germany}
\affiliation{ExtreMe Matter Institute EMMI, GSI Helmholtzzentrum f\"ur Schwerionenforschung GmbH, 64291 Darmstadt, Germany}

\author{V.\ Durant}
\email[Email:~]{durant@theorie.ikp.physik.tu-darmstadt.de}
\affiliation{Institut f\"ur Kernphysik, Technische Universit\"at Darmstadt, 64289 Darmstadt, Germany}
\affiliation{ExtreMe Matter Institute EMMI, GSI Helmholtzzentrum f\"ur Schwerionenforschung GmbH, 64291 Darmstadt, Germany}

\author{J.\ Simonis}
\email[Email:~]{simonis@uni-mainz.de\\
Present address: Institut f\"ur Kernphysik and PRISMA Cluster of
Excellence, Johannes Gutenberg-Universit\"at, 55099 Mainz, Germany}
\affiliation{Institut f\"ur Kernphysik, Technische Universit\"at Darmstadt, 64289 Darmstadt, Germany}
\affiliation{ExtreMe Matter Institute EMMI, GSI Helmholtzzentrum f\"ur Schwerionenforschung GmbH, 64291 Darmstadt, Germany}

\author{A.\ Schwenk}
\email[Email:~]{schwenk@physik.tu-darmstadt.de}
\affiliation{Institut f\"ur Kernphysik, Technische Universit\"at Darmstadt, 64289 Darmstadt, Germany}
\affiliation{ExtreMe Matter Institute EMMI, GSI Helmholtzzentrum f\"ur Schwerionenforschung GmbH, 64291 Darmstadt, Germany}
\affiliation{Max-Planck-Institut f\"ur Kernphysik, Saupfercheckweg 1, 69117 Heidelberg, Germany}

\begin{abstract}
We construct valence-space Hamiltonians for use in shell-model calculations,
where the residual two-body interaction is based on symmetry principles and the
low-momentum expansion from chiral effective field theory. In addition to the usual
free-space contact interactions, we also include novel center-of-mass--dependent
operators that arise due to the Galilean invariance breaking by in-medium effects.
We fitted the low-energy constants to 441 ground- and excited-state energies in
the $sd$ shell and obtained a root-mean-square derivation of 1.8~MeV at leading
order and of 0.5~MeV at next-to-leading order, with natural low-energy constants
in all cases. The developed chiral shell-model interactions enable order-by-order
uncertainty estimates and show promising predictions for neutron-rich isotopes
beyond the fitted data set.
\end{abstract}

\maketitle

\section{Introduction}

The nuclear shell model~\cite{Brow01SM,Caur05RMP,Cora09SM} is a very successful
many-body method, which is widely used for calculations of nuclear-structure properties
in the medium to medium-heavy mass region of the nuclear chart. Typically, the model
space for shell-model calculations includes one major harmonic-oscillator (HO)
shell, or extensions by including
another full shell or some of the lowest-lying subshells. In order to perform calculations
in the nuclear shell model, one requires an effective Hamiltonian that describes the
interactions among nucleons in the valence space under consideration.

There are two common approaches to develop valence-space Hamiltonians, which
typically consist of single-particle energies (SPEs) and two-body matrix elements (TBMEs).
First, there are very successful phenomenological approaches, where an effective
interaction is constructed in a specific valence space by fitting free parameters to
experimental properties in the model space. These are usually theoretically motivated based
on a renormalized realistic interaction, where the TBMEs (or combinations thereof)
and SPEs are then used
to fine-tune the interaction, as in the universal $sd$-shell (USD) interactions of
Ref.~\cite{Brow06USD}. This strategy (see, e.g., Refs.~\cite{Brow01SM,Caur05RMP}
for reviews) typically leads to shell-model interactions that reproduce the experimental data
with a root-mean-square (RMS) deviation of only a few hundred keV. 

Second, valence-space Hamiltonians can be derived using modern ab initio
methods, which can then be used in shell-model calculations. Among those
methods are many-body perturbation theory (MBPT)~\cite{Hjor95MBPT,%
Cora09SM,Holt14Ca,Simo16unc}, the no-core shell model
(NCSM)~\cite{Dikm15NCSMSM,Barr13PPNP}, coupled-cluster theory
(CC)~\cite{Jans14SM,Jans16SM,Hage14RPP}
and the in-medium similarity renormalization group (IM-SRG)~\cite{Tsuk12SM,Bogn14SM,Stro17ENO,Herg16PR}.
All of these methods start from a few-body Hamiltonian, which typically
consists of two- and three-body interactions from chiral effective field theory (EFT).
These methods do not achieve the same overall accuracy as the
phenomenological fits, but they can provide uncertainty estimates.

In this paper, we use chiral EFT as a general operator basis at low energies
and its capability to estimate theoretical uncertainties due to the EFT expansion
to develop chiral shell-model interactions, where the low-energy couplings
(LECs) are fit directly to data in the $sd$ shell. Chiral EFT provides a systematic
expansion of strong interactions at low energies based on general symmetry
principles in terms of nucleon and pion degrees of freedom~\cite{Epel09RMP,Mach11PR}.
Following Weinberg's power counting~\cite{Wein90NFch,Wein91chNp},
chiral EFT predicts a hierarchy of two- and many-body interactions governed
by an expansion in powers of $(Q/\Lambda_b)^\nu$ with order $\nu \geqslant 0$,
were $Q$ is a generic low-momentum scale or the pion mass $m_\pi$ and 
$\Lambda_b \sim 500$~MeV is the breakdown scale of the EFT. Chiral EFT
includes two-nucleon interactions at leading order (LO, $Q^0$) and many-body
interactions start at next-to-next-to-leading order (N$^2$LO, $Q^3$).

Because the pion-exchange interactions describe long-range physics, which is
not renormalized in the medium, we take the long-range pion-exchange contributions
directly as in free-space nuclear forces~\cite{Epel09RMP,Mach11PR}. The
short-range contact interactions encode physics beyond the degrees of freedom
resolved in the EFT and therefore, for chiral shell-model interactions we fit these
directly to data in the $sd$ shell. However, in the valence space, the presence
of the core breaks Galilean invariance, and therefore novel short-range operators
are possible that depend on the two-body center-of-mass (CM) momentum
(or on the CM orbital angular momentum). These have been explored in the
context of Fermi liquid theory in Ref.~\cite{Schw03noncentral} and include
operators that are known as antisymmetric spin-orbit interactions in the context
of the shell model (see, e.g., Ref.~\cite{Conz73ALS}). They enter at next-to-leading
order (NLO, $Q^2$) in Weinberg counting, and we explore them for the
first time in shell-model interactions.

In this work, we construct valence-space Hamiltonians in the $sd$ shell based 
on chiral EFT operators up to NLO. We fit the LECs to
441 ground- and excited-state energies in this model space. The LECs absorb
in-medium effects due to the truncation of the model space. We will show the
significance of the novel CM-dependent operators by constructing a full 
valence-space (vs) NLO interaction, which we label NLO$_\mathrm{vs}$, and
comparing it to results for a NLO interaction that uses only free-space
operators from chiral EFT. We also compare our chiral shell-model interactions
with the USD interactions from Ref.~\cite{Brow06USD}. Moreover, we explore order-by-order
uncertainty estimates and show promising predictions for neutron-rich isotopes
beyond the fitted data set.

This paper is organized as follows: In Sec.~\ref{sec:ops}, we discuss the
free-space contact interactions and introduce the new CM-dependent operators
at NLO. The partial-wave decomposition of the operators is given in
App.~\ref{sec:apppwd}. The second part of Sec.~\ref{sec:ops} discusses
the transformation to TBMEs in a HO basis and regulator aspects. Details
on the transformation are given in App.~\ref{sec:detailsho}. We discuss specifics
on the fitting process and give an overview of the quality of our fits in Sec.~\ref{sec:fit}.
In Sec.~\ref{sec:res}, we show our results and predictions for ground-state energies
and spectra, including estimates of the theoretical uncertainties. Finally,
we summarize and give an outlook in Sec.~\ref{sec:sum}.

\section{Valence-shell interactions}
\label{sec:ops}

\subsection{Operators from chiral EFT}

Following Weinberg's power counting~\cite{Wein90NFch,Wein91chNp}, there are two
LECs at LO and seven new LECs at NLO. The LO and NLO contact interactions have
the following form in momentum space:
\begin{equation}
\braket{\vec{p}| \o{V}_{\text{cont}}^{(\mathrm{LO})}| \vec{p}'} 
= C_S + C_T  \, \vec{\sigma}_1\cdot\vec{\sigma}_2 \,,
\end{equation}
and
\begin{align}
\braket{\vec{p}| \o{V}_{\text{cont}}^{(\mathrm{NLO})}|\vec{p}'} &= C_1 \vec{q}^2 + C_2 \vec{k}^2 
+ \left(C_3 \vec{q}^2 + C_4 \vec{k}^2\right) \vec{\sigma}_1\cdot\vec{\sigma}_2  \nonumber\\ 
&+ C_5 \, \frac{\ci}{2}\left(\vec{\sigma}_1+\vec{\sigma}_2\right)\cdot\left(\vec{q}\times\vec{k}\right) 
\nonumber\\ 
&+ C_6 \left(\vec{\sigma}_1 \cdot\vec{q}\right)\left(\vec{\sigma}_2 \cdot\vec{q}\right)
+ C_7 \left(\vec{\sigma}_1 \cdot\vec{k}\right)\left(\vec{\sigma}_2 \cdot\vec{k}\right) ,
\end{align}
where $\vec{p}$ and $\vec{p}'$ are the final and initial relative momenta
with $\vec{p} = (\vec{p}_1 - \vec{p}_2)/2$, $\vec{q}$ is the momentum
transfer $\vec{q}=\vec{p}-\vec{p}'$ and $\vec{k}$ is the average momentum
$\vec{k}=(\vec{p}+\vec{p}')/2$. The partial-wave decomposition of the free-space
contact interactions is given in App.~\ref{sec:freepwd}.

The additional operators in the valence space,
due to broken Galilean invariance by the presence of the core, depend explicitly
on the two-body CM momentum $\vec{P} = \vec{p}_1 + \vec{p}_2$.
We count powers of $\vec{P}$ as powers of $Q$, as they are set by the same
scale (the inverse oscillator length) in a shell-model basis. Thus, the first contributions
from these operators arise at NLO. We label the CM-dependent part of the
contact interactions as NLO$_\mathrm{vs}$, where vs is short for valence space.
These take the following form in momentum space:
\begin{align}
\braket{\vec{p},\vec{P}| \o{V}_{\text{cont}}^{(\mathrm{NLO}_{\mathrm{vs}})}|\vec{p}',\vec{P}} 
&= P_1 \vec{P}^2  + P_2 \vec{P}^2  \vec{\sigma}_1\cdot\vec{\sigma}_2  \nonumber\\ 
&+ P_3 \, \ci \left(\vec{\sigma}_1-\vec{\sigma}_2\right)\cdot\left(\vec{q}\times\vec{P}\right)\nonumber\\
&+ P_4 \left(\vec{\sigma}_1\times\vec{\sigma}_2\right)\cdot\left(\vec{k}\times\vec{P}\right) \nonumber\\
&+ P_5 \left(\vec{\sigma}_1 \cdot\vec{P}\right)\left(\vec{\sigma}_2 \cdot\vec{P}\right)\,.
\end{align}
The CM-dependent interactions include central parts, given by the LECs
$P_1$ and $P_2$, the difference- and cross-vector operators determined by $P_3$
and $P_4$, and a CM tensor operator, given by $P_5$. The latter three have been introduced
and discussed in the context of noncentral interactions in Fermi liquid
theory~\cite{Schw03noncentral}. As shown by the partial-wave decomposition
in App.~\ref{sec:cmpwd}, the central and tensor parts are diagonal in two-body
spin $s$, relative orbital angular momentum $l$, and total (relative plus spin) angular momentum
$j$, and they only contribute to the relative $^1S_0$ and $^3S_1$ waves.
Note that in the presence of local regulators, regulator artifacts would also
lead to contributions in higher partial waves (see, e.g., Ref.~\cite{Huth17Fierz}).
Moreover, the central parts are diagonal in CM angular momentum $L$.

The difference- and cross-vector operators are spin-violating~\cite{Schw03noncentral}
and mix spin-singlet $^1S_0$ ($^1P_1$) with spin-triplet $^3P_j$ ($^3S_1$)
relative partial waves. At NLO$_\mathrm{vs}$, they do not contribute to higher $l$
waves. As a result of the $S$-$P$ mixing and parity conservation, the spin-violating
interactions also change
the CM angular momentum $L,L'$ and are not necessarily diagonal in $j,j'$.
In the shell-model context, their structure is similar to the anti-symmetric spin-orbit
interaction (see, e.g., Ref.~\cite{Conz73ALS}).

In order to investigate the impact of the different CM-dependent interactions, we
use in the following the notation NLO$_{\mathrm{vs}_{c,v,t}}$ when only central,
only vector, or only tensor operators are included, respectively.

\subsection{Transformation to HO basis and regulators}

In order to apply the momentum-space interactions in the valence space, we
transform them to antisymmetrized, normalized two-body HO states.
As detailed in App.~\ref{sec:detailsho}, this leads to TBMEs of the form
\begin{equation}
\braket{(n_1 l_1 j_1)(n_2 l_2 j_2) J T | V | (n'_1 l'_1 j'_1)(n'_2 l'_2 j'_2) J T} ,
\end{equation}
where $(n_i, l_i, j_i)$ are the single-particle radial, orbital angular momentum,
and total angular momentum quantum numbers, and $J,T$ are the
two-body total angular momentum and isospin, respectively.

\begin{figure}[t]
\includegraphics[width=\columnwidth]{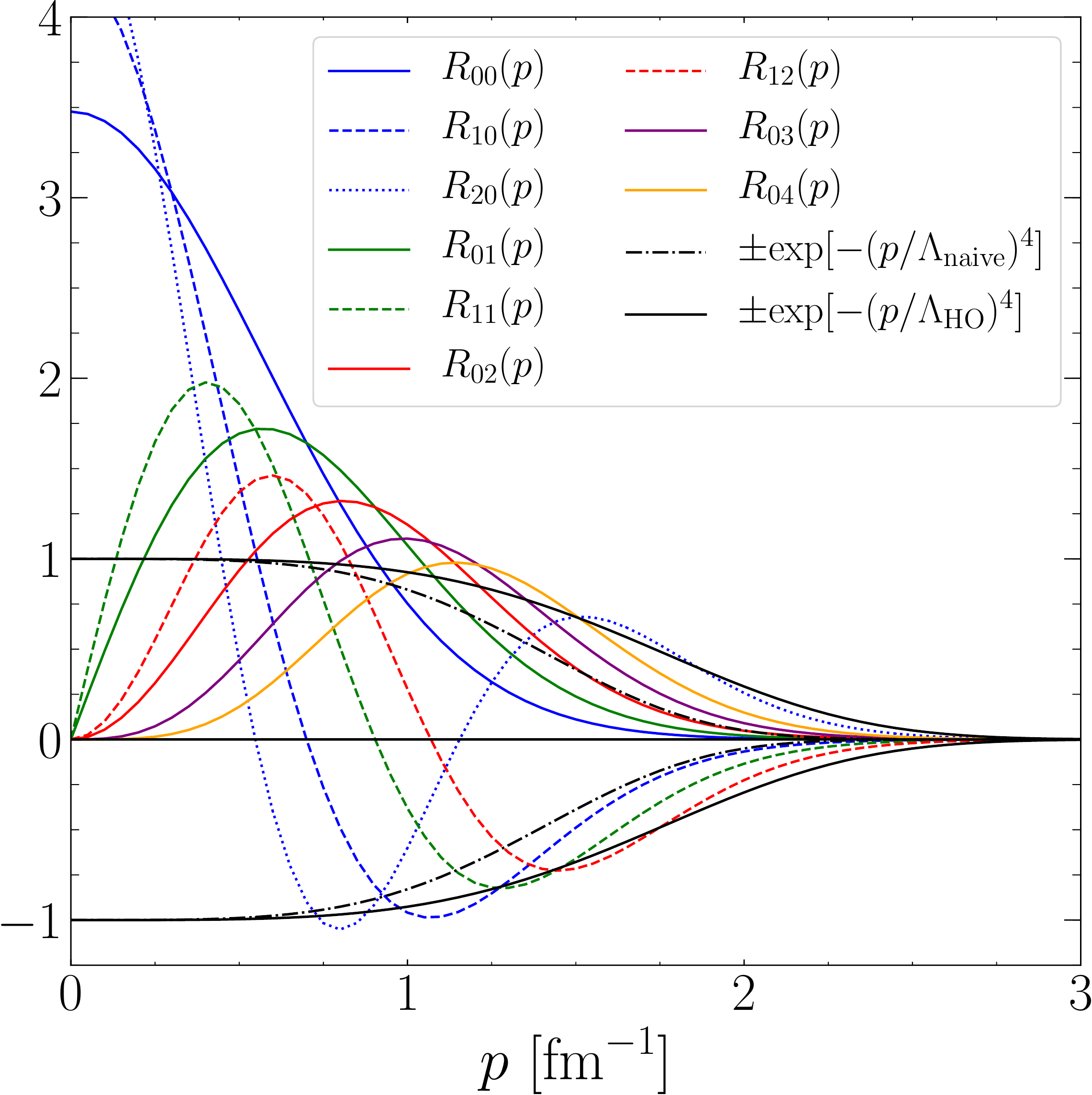}
\caption{Radial wave functions relevant for $sd$-shell TBMEs in
comparison to typical regulators used in chiral EFT with the naive cutoff
estimate $\Lambda_\mathrm{naive} = 300\,\mathrm{MeV}$ and
$\Lambda_\mathrm{HO} = 375\,\mathrm{MeV}$. See text for details.}
\label{fig:cutoff}
\end{figure}

The radial HO wave functions are given by
\begin{equation}
R_{nl}(p) = N_{nl} \, (pb)^l \, \exp\Bigl[-\frac{1}{2} (pb)^{2} \Bigr]  L_n^{l+\frac{1}{2}}\Bigl((pb)^2\Bigr) \,,
\end{equation}
and are plotted in Fig.~\ref{fig:cutoff} for different $n,l$ quantum numbers
relevant for $sd$-shell TBMEs. The oscillator
length $b=\sqrt{\hbar/(m \omega)}$ is used with $\hbar \omega = 13.53$~MeV to reproduce
the radius of the $^{16}$O core. For completeness, the normalization is $N_{nl} = b^{3/2}
\sqrt{2 n!/\Gamma(n+l+3/2)}$ and $L_n^{l+\frac{1}{2}}$ are generalized Laguerre polynomials.

Figure~\ref{fig:cutoff} shows that the radial wave functions involved in a limited valence
space automatically cut off the high-momentum parts, and therefore no additional
momentum-space regulator functions are necessary.
In fact, one can naively estimate the
cutoff in energy due to the basis truncation by
\begin{equation}
\frac{\Lambda_\mathrm{naive}^2}{m_N} \sim E \leqslant \varepsilon_1 + \varepsilon_2 
=  2 (N_\mathrm{valence} + 3/2) \hbar \omega \,.
\end{equation}
For the $sd$ shell it follows that $\Lambda_\mathrm{naive} \approx 300\,\mathrm{MeV}$.
A more sophisticated estimate is given in Ref.~\cite{Koni14UVextrapol} leading to a cutoff
estimate for the $sd$ shell $\Lambda_\mathrm{HO} \approx 375\,\mathrm{MeV}$.
In Fig.~\ref{fig:cutoff}, we also compare the radial wave functions relevant for $sd$-shell
TBMEs with commonly used regulators from chiral EFT with the two cutoff estimates
described above. We observe that the radial wave functions indeed have a similar behavior
in the high-momentum part as the regulator function with $\Lambda_\mathrm{HO} = 
375\,\mathrm{MeV}$. Hence, there is no necessity for additional momentum-space regulator
functions for the contact interactions.

\section{Fit}
\label{sec:fit}

\subsection{Data set}

\begin{figure}[t]
\includegraphics[width=\columnwidth]{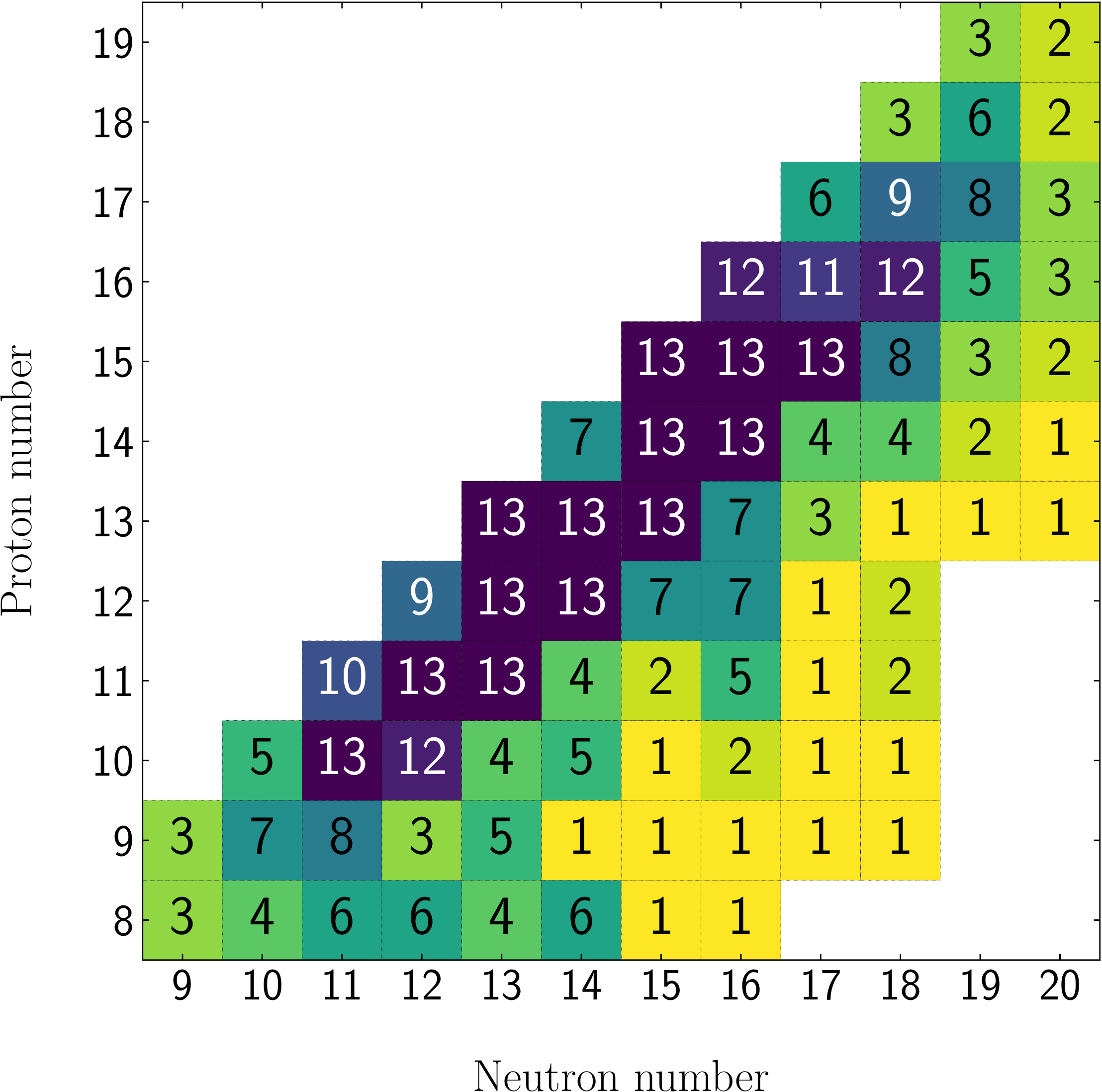}
\caption{Graphical representation of the 441 experimental states used in our
$sd$-shell fits. Each square shows the number of states fitted for a given
isotope, where the color coding gets darker with increasing number of states.}
\label{fig:set}
\end{figure}

With the LO and different NLO operators in place, the next step is to determine the LECs
by fits to experimental data. For our data set, we consider 441 out of the 608 states that
were used for the USDA/USDB fits~\cite{Brow06USD}. This set is smaller than the one used
for USDA/USDB, because we included at most 12 excited states for a given isotope.
The number of states we fit to each
isotope is visualized in Fig.~\ref{fig:set}. As in Ref.~\cite{Brow06USD}, we apply a proton
number dependent Coulomb correction to the experimental ground-state energies, 
so that we can focus on the strong interaction part. The Coulomb corrections used are
listed in Table~\ref{tab:coulcorr}.

\begin{table}[t]
{\renewcommand{\arraystretch}{1.3}
\begin{ruledtabular}
\centering
\caption{Coulomb correction for a given proton number $Z$ from Ref.~\cite{Brow06USD}.
We have corrected the experimental ground-state energies by subtracting the Coulomb
correction.}
\label{tab:coulcorr}
\begin{tabular}{cc|c}
$Z$ & element & Coulomb correction [MeV] \\ 
\hline 
\hphantom{0}8  & O\hphantom{e} & \hphantom{0}0.00 \\ 
\hphantom{0}9 & F\hphantom{e}  & \hphantom{0}3.48  \\ 
10 & Ne & \hphantom{0}7.45 \\ 
11 & Na & 11.73 \\
12 & Mg & 16.47 \\
13 & Al & 21.48 \\
14 & Si & 26.78 \\
15 & P\hphantom{e} & 32.47 \\
16 & S\hphantom{e} & 38.46 \\
17 & Cl & 44.74 \\
18 & Ar & 51.31 \\
19 & K\hphantom{e} & 58.14 \\
\end{tabular}
\end{ruledtabular}} 
\end{table}

\subsection{Optimization}

As mentioned above, the HO frequency is set by the $^{16}$O radius. For different
mass number $A$, we apply a scaling factor $(18/A)^{1/3}$ to the TBMEs to correct
the HO frequency for larger nuclei, which is a standard procedure in shell-model
calculations. Our SPEs are set to reproduce the one-neutron separation energy
and the first two excited states of $^{17}$O. The LECs of the contact operators are
then determined by a $\chi^2$ minimization. The $\chi^2$ value per datum is calculated
as follows:
\begin{equation}
\chi^2=\frac{1}{N-p}\sum\limits_{i=1}^{N}\left(\frac{E^{\text{exp}}_i-E^{\text{th}}_i}{\sigma_i}\right)^2 ,
\end{equation}
where $N$ is the total number of states and $p$ the number of parameters (LECs)
in the fit. The experimental energy $E^{\text{exp}}_i$ is taken from the data set mentioned
above, and the theoretical result $E^{\text{th}}_i$ is obtained by diagonalizing the
valence-space Hamiltonian. For this, we use the shell-model code
ANTOINE~\cite{Caur99antoine,Caur05RMP}. The uncertainty $\sigma_i$  is given by
$\sigma_i^2=(\sigma_{i}^{\text{exp}})^2+ (\sigma_i^{\text{th}})^2$,
where we take the experimental uncertainty from the data set and for the
theoretical uncertainty we use a constant value $\sigma_i^{\text{th}}=0.1\,\mathrm{MeV}$
as in Ref.~\cite{Brow06USD}. In future work, we will also propagate the uncertainty
from the EFT expansion, which we explore here first after the fits in Sec.~\ref{sec:uq}.
For the optimization, we use the linear combination method, described in 
Ref.~\cite{Brow06USD}. The routine shows a fast and stable
convergence, but requires a linear dependence on our LECs, which rules out
uncertainty estimates that explicitly depend on the parameters. We have also checked
that the fit is stable under further optimization with POUNDerS algorithm~\cite{POUNDerS,Wild15POUNDerS} or
using the Nelder-Mead method~\cite{NM1965}. Finally, we have considered
several starting points for the fits: all LECs set to zero; starting from LECs fit to 
reproduce the USDA/B interactions; and starting from LECs fit to reproduce MBPT
TBME from chiral NN+3N interactions. We have observed that the fits
based on these starting points all lead to the same minimum.

As our theoretical uncertainty has no statistical interpretation, neither does
the resulting $\chi^2$ value, and thus, we rather compare the RMS
deviation to experiment for different interactions. The RMS deviation is
given by
\begin{equation}
\text{RMS} = \sqrt{\frac{1}{N} \sum\limits_{i=1}^{N} \left(E^{\text{exp}}_i-E^{\text{th}}_i
\right)^2} \,.
\end{equation}

\subsection{Overview of comparison with experiment}

\begin{figure*}[p]
\begin{turn}{90}
\begin{minipage}{1.41\textwidth}
\centering
\includegraphics[width=\textwidth]{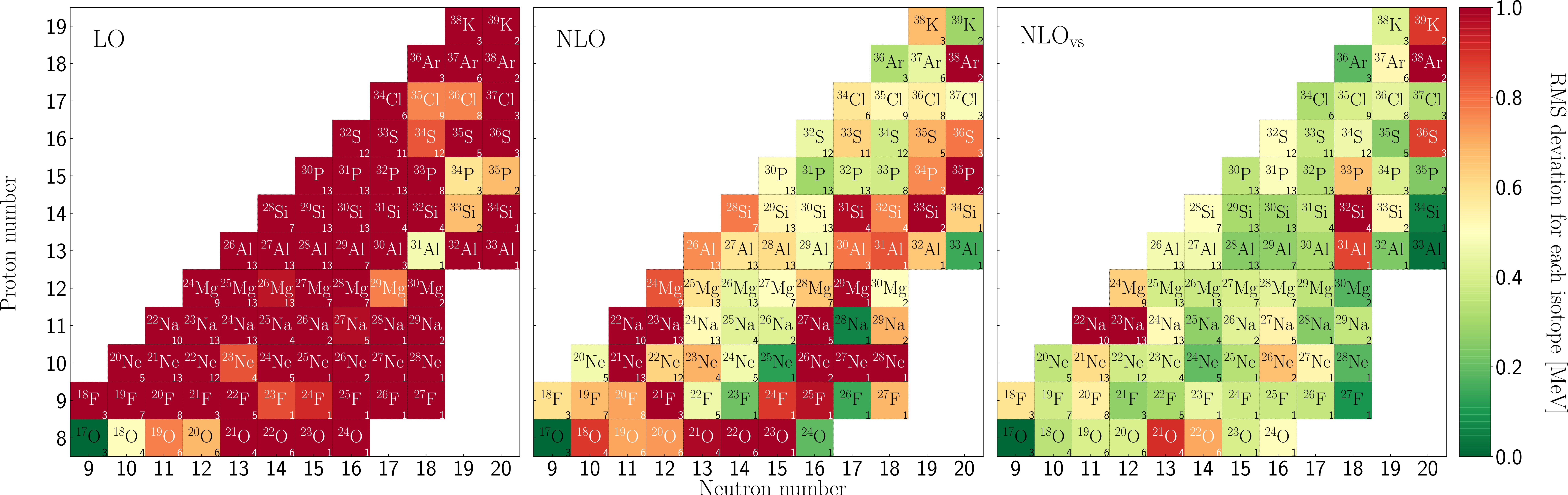}
\caption{Graphical representation of the RMS deviation from experiment for each
fitted nucleus in the $sd$ shell. The figure shows the results for the chiral shell-model interactions
at LO (left), NLO (middle), and NLO$_\mathrm{vs}$(right). The color coding of the
RMS deviation is given in the bar on the right. Isotopes with a small RMS deviation
are colored green, while those with a large deviation are colored red. Each square
shows the isotope label and the number of fitted states in the bottom right corner.
The text color changes from black to white for RMS deviations larger than 0.7 MeV.}
\label{fig:singlenucrms}
\end{minipage}
\end{turn}
\end{figure*}

\begin{figure*}[p]
\begin{turn}{90}
\begin{minipage}{1.41\textwidth}
\includegraphics[width=\textwidth]{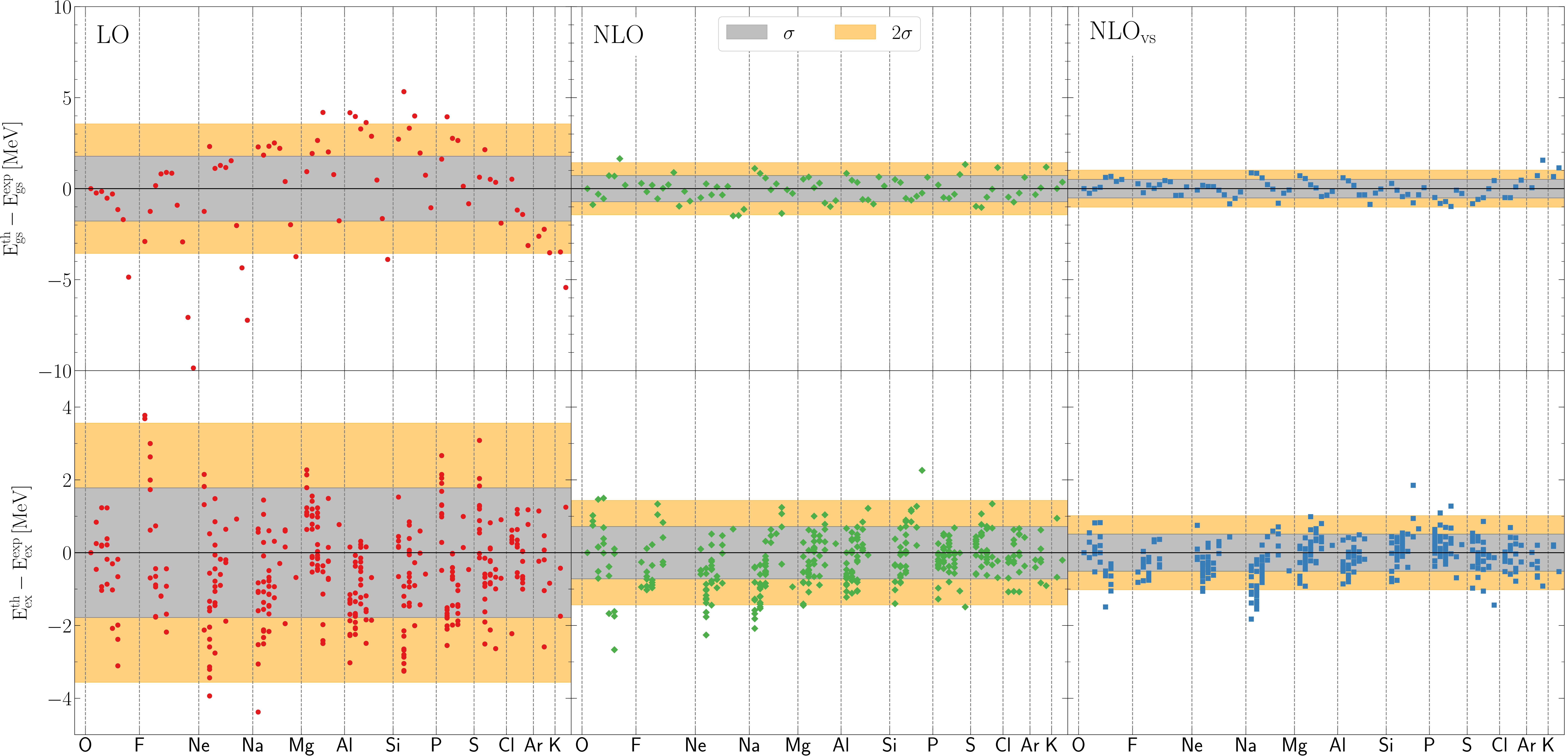}
\caption{Energy differences of the result from the chiral shell-model interactions
and experiment in MeV. The upper row shows results for ground-state (gs) energies
whereas the lower row is for the energies of excited (ex) states. Each dot represents
a single state. The dots are ordered from oxygen to potassium, and within each
bin they are ordered according to their mass number. From left to right, we show
results for the LO, NLO, and NLO$_\mathrm{vs}$ interaction. The gray (orange)
bands show the statistical $\sigma$ ($2\sigma$) spread, given by the RMS deviation.}
\label{fig:cmp_all}
\end{minipage}
\end{turn}
\end{figure*}

In Fig.~\ref{fig:singlenucrms} we show the RMS deviation from experiment for each
fitted nucleus in the $sd$ shell for the chiral shell-model interactions at LO (left), NLO
(middle), and NLO$_\mathrm{vs}$ (right). The RMS deviation is given by a color 
coding that ranges from 0~MeV (green) to 1~MeV (red). The results show a striking
improvement from LO to NLO and a further improvement from NLO to NLO$_\mathrm{vs}$,
where at NLO$_\mathrm{vs}$, there are only a few outliers with large RMS deviations.
This demonstrates the impact of the new CM-dependent operators.

We also show a quantitative overview of the comparison with experiment in
Fig.~\ref{fig:cmp_all}. The figure is divided into two rows, where the upper row
shows the difference between theoretical and experimental ground-state energies
and the lower row is for the difference between theoretical and experimental
excitation energies. The columns show again the results for the LO (left), 
NLO (middle), and NLO$_{\mathrm{vs}}$ (right) shell-model interactions. 
The gray (orange) bands show the $\sigma$ ($2\sigma$) intervals given by the
RMS deviation. The order-by-order improvement from LO to NLO and from NLO to 
NLO$_{\mathrm{vs}}$, already seen globally in Fig.~\ref{fig:singlenucrms}, is
clearly visible from the decreasing $\sigma$ bands from left to right and from
the systematically decreasing individual energy differences. Overall, we observe
a very good reproduction of experiment at NLO$_\mathrm{vs}$.

The results for the ground-state energies at LO in Fig.~\ref{fig:cmp_all}
show a systematic deviation from
experiment with increasing neutron richness, especially for the oxygen to silicon
isotopes, where the LO shell-model interaction leads to overbound states with
respect to experiment. This trend seems to be resolved at NLO, where no clear
pattern is visible. However, at NLO$_{\mathrm{vs}}$, there is again a deficiency
in the isospin dependence for the neon to aluminum isotopic chains. It will be interesting
to see whether this will be improved at N$^2$LO, and whether this can be traced
back to the inclusion of three-nucleon forces~\cite{Otsu10Ox}, which enter at N$^2$LO.

Systematic trends of this type are not visible in the energy differences for the
excited states in Fig.~\ref{fig:cmp_all}. Note that the number of excited states is higher
for nuclei close to stability (see also Fig.~\ref{fig:set}), so that there are more
points shown at the beginning of each element in Fig.~\ref{fig:cmp_all}.
However, it stands out that there is little to no improvement in the first two sodium
isotopes ($^{22}$Na and $^{23}$Na) from NLO to NLO$_{\mathrm{vs}}$, which
exemplary shows that additional operator structures are necessary to reach higher
accuracies in the fit.

\subsection{Comparison to USD-type interactions}

\begin{figure}[t]
\includegraphics[width=\columnwidth]{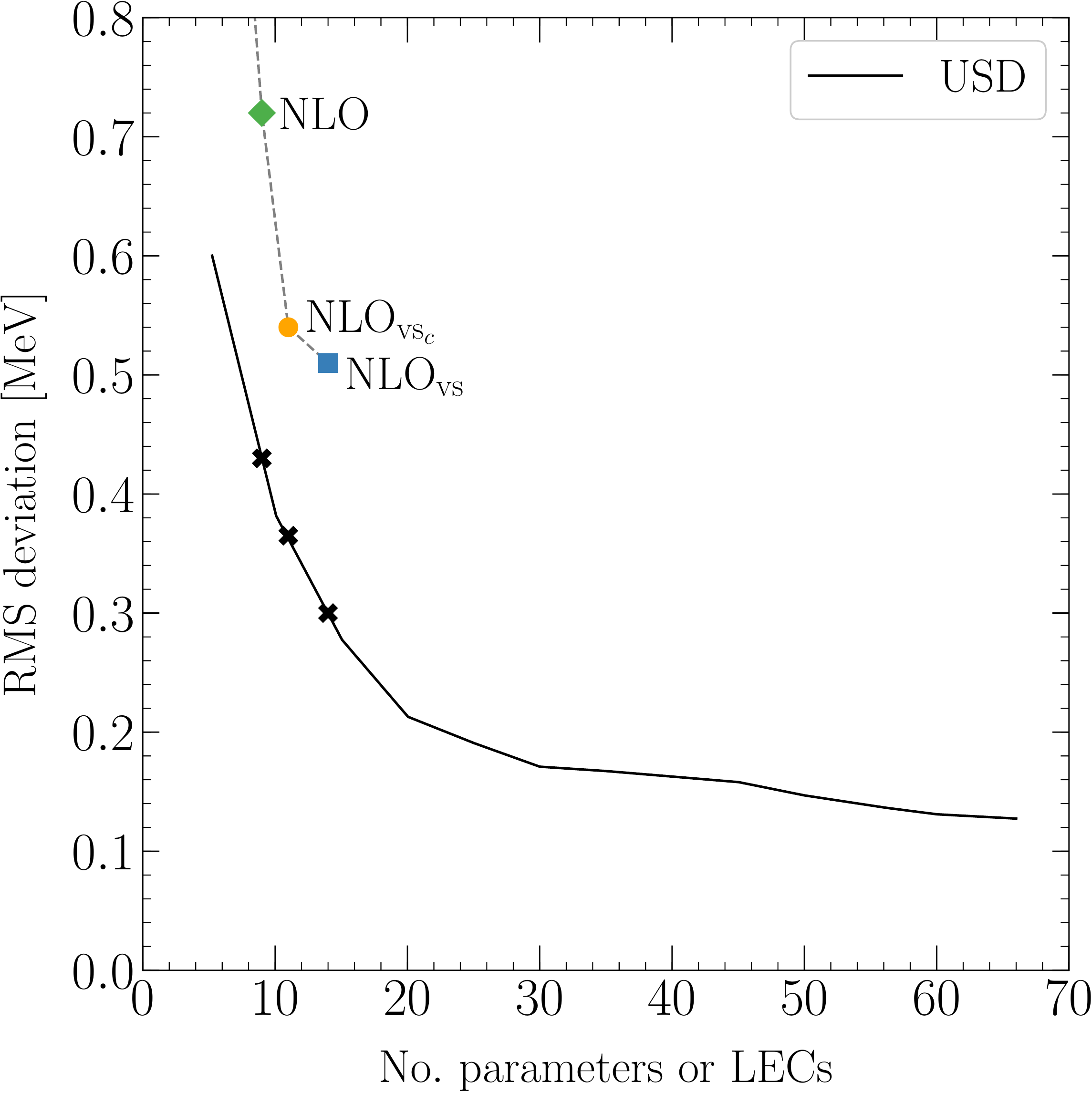}
\caption{Root-mean-square deviation of the chiral and USD-type interactions as a function
of the number of parameters. The results for the USD fit are taken from Ref.~\cite{Brow06USD}.
The figure shows the free-space operator NLO interaction (green diamond),
the NLO$_\mathrm{vs_\mathrm{c}}$ interaction including only the central CM-dependent
operators in addition to NLO (yellow circle), and the full
NLO$_\mathrm{vs}$ shell-model interactions (blue square).}
\label{fig:usdcmp}
\end{figure}

In addition to the direct comparison with experiment, we study how the developed
chiral shell-model interactions perform compared to USD-type interactions, where
the TBMEs are not determined by a basis of operators but are fit overall.
The RMS deviation of the USD fit is taken from Fig.~4 of Ref.~\cite{Brow06USD} and
is shown as solid line as a function of the number of parameters in Fig.~\ref{fig:usdcmp}.
Note that the USD fit was to a data set with 608 states, while our results are for the data
set of 441 states described above, so the comparison is not completely one-to-one.

In Fig.~\ref{fig:usdcmp}, we plot the RMS deviation as a function of the number
of LECs for the different chiral shell-model interactions developed in this work. In order to
assess the impact of the new CM-dependent operators, we also analyze the 
central (vs$_c$), vector (vs$_v$), and tensor (vs$_t$) contributions separately. The
RMS deviations for all interactions and the number of LECs are given in
Table~\ref{tab:rms}. Note that in comparison to the RMS deviation for a given
nucleus (see Fig.~\ref{fig:singlenucrms}) or for ground- and excited states separately
(see Fig.~\ref{fig:cmp_all}) the RMS deviations discussed here are with respect to the
full data set considered. As shown in Table~\ref{tab:rms}, the RMS deviation improves
1.8~MeV at LO to 0.7~MeV at NLO and 0.5~MeV at NLO$_{\mathrm{vs}}$.

To guide the comparison with the USD-type interactions in Fig.~\ref{fig:usdcmp}, the
latter are marked by a cross for 9, 11, and 14 parameters, which corresponds to the
same number of LECs as the NLO, NLO$_{\mathrm{vs}_{c}}$ (or NLO$_{\mathrm{vs}_{v}}$),
and NLO$_{\mathrm{vs}}$ interactions, respectively (see Table~\ref{tab:rms}).
Recall that the USDA (USDB) interactions correspond to the USD fit with 30 (56)
parameters~\cite{Brow06USD}. We find a similar rapid decrease of the RMS deviation
with increasing number of LECs, although for the same number of parameters
the optimal USD fit has $\sim 200$~keV smaller RMS deviation. Moreover, we show in
Fig.~\ref{fig:usdcmp} explicitly the NLO$_{\mathrm{vs}_{c}}$ result, because the
central CM-dependent operators constitute the largest source of improvement
compared to considering only free-space operators (see also Tab.~\ref{tab:rms}).

\begin{table}[t]
{\renewcommand{\arraystretch}{1.3}
\begin{ruledtabular}
\caption{Number of fitted LECs for the different chiral shell-model interactions considered
in this work. The first two rows show the LO and NLO interactions based on free-space
operators. The following rows show NLO interactions that include the CM-dependent
operators from Sec.~\ref{sec:ops}. To distinguish between central ($c$),
vector ($v$), and tensor ($t$) contributions, we label them vs$_c$, vs$_v$, and vs$_t$,
respectively. The full valence-space interaction
in the last row is labeled NLO$_\mathrm{vs}$. We give the RMS deviation from
experiment for these fitted interactions and compare them to the RMS deviation
of the USD fit from Ref.~\cite{Brow06USD} for the same number of parameters. 
The rows are ordered with increasing number of LECs and decreasing RMS deviation.}
\label{tab:rms}
\begin{tabular}{l|c|c|c}
Interaction & $\#$LECs & RMS [keV] & USD [keV] \\ 
\hline 
LO & \hphantom{0}2 & 1780 & $-$ \\ 
NLO & \hphantom{0}9 & \hphantom{0}718 & 430 \\ 
NLO$_{\mathrm{vs}_{t}}$ & 10 & \hphantom{0}641 & 380 \\ 
NLO$_{\mathrm{vs}_{v}}$ & 11 & \hphantom{0}678 & 370 \\ 
NLO$_{\mathrm{vs}_{c}}$ & 11 & \hphantom{0}538 & 370 \\ 
NLO$_{\mathrm{vs}}$ & 14 & \hphantom{0}510 & 300 \\ 
\end{tabular} 
\end{ruledtabular}} 
\end{table}

\subsection{Monopole matrix elements and\\ low-energy constants}

The monopole matrix elements play a special role in the shell model and for shell
structure~\cite{Caur05RMP,Otsu10Ox,Otsu10VMU,Holt12Ca}. They determine
the energy gaps between the single-particle orbitals, leading to effective SPEs.
Using a short-hand notation for the TBMEs, $\braket{abJT | V | cdJT}$, where
the combined index $i$ is short for $(n_i l_i j_i)$, the monopole matrix elements are
obtained by angle averaging, i.e., by a weighted average over all possible values
of the total angular momentum,
\begin{equation}
V_{ab}^T= \frac{\sum\limits_J (2J+1) \braket{abJT|V|abJT}}{\sum\limits_J (2J+1)} \,,
\end{equation}
where in the $sd$ shell $a,b$ consists of the $0d_{5/2}$, $0d_{3/2}$, and $1s_{1/2}$
orbitals, which are uniquely labeled by twice their total angular momentum label (i.e., 5, 3, and 1).

\begin{figure*}[t!]
\centering
\includegraphics[width=\textwidth]{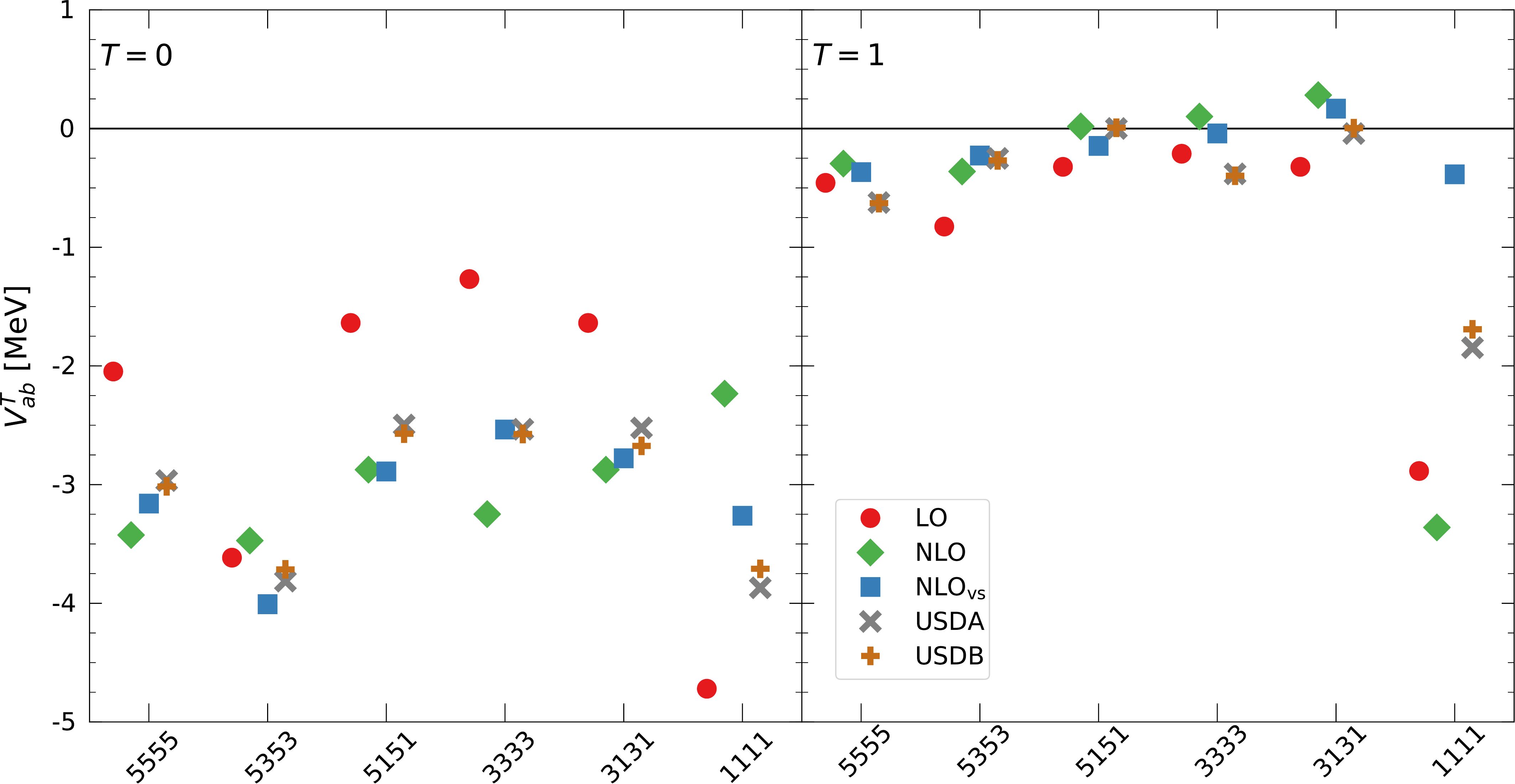}
\caption{Monopole matrix elements of the different chiral shell-model interactions for mass
number $A=18$. The left panel shows the monopole matrix elements for isospin $T=0$
and the right panel for $T=1$. The chiral shell-model interactions at LO, NLO, and NLO$_\mathrm{vs}$
are shown together with the monopole matrix elements of the USDA and USDB
interactions from Ref.~\cite{Brow06USD}.
The matrix elements are labeled by $2j_a\,2j_b\,2j_a\,2j_b$.}
\label{fig:monopoles}
\end{figure*}

\begin{figure*}[t!]
\centering
\includegraphics[width=\textwidth]{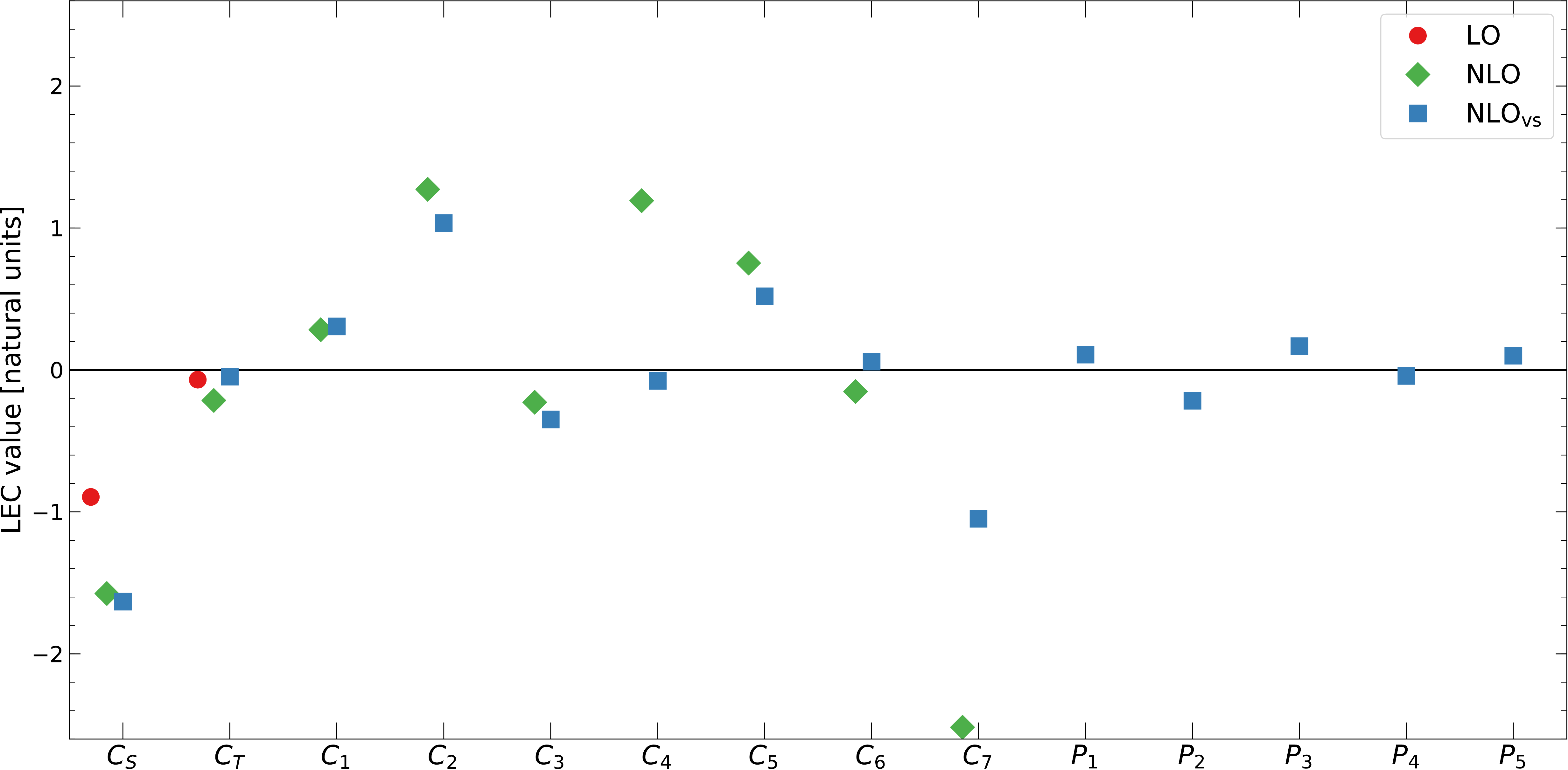}
\caption{Fitted LECs at LO, NLO, and NLO$_\mathrm{vs}$ in natural units, which
are obtained using Eqs. \eqref{eq:natLO}--\eqref{eq:natNLO}.}
\label{fig:lec}
\end{figure*}

Figure~\ref{fig:monopoles} shows the monopole matrix elements of the chiral
shell-model interactions at LO, NLO, and NLO$_{\mathrm{vs}}$ as well as those
of the USDA and USDB interaction for $A=18$ (i.e., without applying the scaling
with $\hbar \omega$). In the $T=0$ channel (left panel of Fig.~\ref{fig:monopoles}),
the monopole matrix elements at LO (except for 5353) deviate significantly from
the other interactions, while at NLO and especially at NLO$_{\mathrm{vs}}$ they
are similar to the monopole matrix elements of USDA/USDB. In general the change
from NLO to NLO$_{\mathrm{vs}}$ is small except for the higher lying 1111
and 3333 orbitals. For the $T=1$ channel (right panel of Fig.~\ref{fig:monopoles}), the
changes from LO to NLO are significantly smaller, and there are only notable 
deviations from USDA/USDB for the 1111 monopole matrix element.
The latter was also observed in microscopic calculations of valence-space
Hamiltonians~\cite{Otsu10Ox}.

The resulting LECs at different orders are shown in Fig.~\ref{fig:lec}. To express them in
natural units (see, e.g., Ref.~\cite{Epel05EGMN3LO}), we multiply the LO and NLO LECs by
\begin{align}
C_\text{LO} [\text{nat. units}] &= C_\text{LO} \cdot F_\pi^2 \,, \label{eq:natLO} \\
C_\text{NLO} [\text{nat. units}] &= C_\text{NLO} \cdot \left(F_\pi \Lambda\right)^2 \,, \\
P_i [\text{nat. units}] &= P_i \cdot \left(F_\pi \Lambda\right)^2 \,, \label{eq:natNLO}
\end{align}
with $\Lambda = \Lambda_\text{HO} = 375$~MeV and pion decay constant $F_\pi = 92.4$~MeV.
As shown in Fig.~\ref{fig:lec}, all fitted LECs at all orders come out to be natural, or are very small
in some cases. Wigner symmetry given by $C_S \gg C_T$ is also fulfilled by our interactions.
Note that neither naturalness nor Wigner symmetry was imposed as a constraint on the fit.
The LECs of the new CM-dependent operators are given
by $P_{1-5}$ in Fig.~\ref{fig:lec}. We find that all $P_i$ are similar in magnitude. 
Finally, the changes from LO to NLO and NLO$_\mathrm{vs}$ are also
systematic for the LECs, with larger changes from NLO to NLO$_\mathrm{vs}$ mainly for
$C_4$ and $C_7$.

\section{Results}
\label{sec:res}

After a discussion of our fits and the overview of the comparison to experiment
and to USD-type interactions in the previous section, we next present a more
detailed picture of the quality of the chiral shell-model interactions. In most cases
the experimental data shown are from the atomic mass evaluation~\cite{Wang12AME12}
for the ground-state energies and from Ref.~\cite{nndc14ENSDF} for excitation energies,
otherwise the experimental reference is given explicitly. Moreover, experimental states
included in the fit are shown in gray, and in red for predictions. We also
provide the TBMEs and SPEs of the NLO$_\mathrm{vs}$ interaction in App.~\ref{sec:tbmes}.

\subsection{Uncertainty estimates}
\label{sec:uq}

The EFT enables estimates of the theoretical uncertainty due to the truncation of the
chiral expansion. We explore these uncertainties here after the chiral shell-model
interactions have been fit, but will explore the fits within the optimization in future
work. The purpose of the present uncertainty study is to obtain a feeling for these in
the context of the shell-model calculations. We emphasize that these theoretical
uncertainties do not include the systematic uncertainties from the shell-model basis
or from possible states that have a small overlap with $sd$-shell configurations.

For the ground-state energies, we directly apply the chiral EFT uncertainty estimate
from Ref.~\cite{Epel15improved} and show the resulting uncertainties in 
Fig.~\ref{fig:nuc_chains}. These are obtained at LO and NLO using
\begin{align}
\Delta E_{\text{LO}}^{\text{gs}} &= \abs{E_{\text{LO}}^{\text{gs}}} Q^2 \,, \label{eq:errgslo} \\[1mm]
\Delta E_{\text{NLO}}^{\text{gs}} &= \max\bigl(\abs{E_{\text{LO}}^{\text{gs}}} Q^3,\,
\abs{E_{\text{LO}}^{\text{gs}}-E_{\text{NLO}}^{\text{gs}}}Q \bigr) \,, \label{eq:errgsnlo}
\end{align}
where $Q=m_\pi/\Lambda_b$ with pion mass $m_\pi$, and 
we take $\Lambda_b = \Lambda_\mathrm{HO} = 375$~MeV.

For excitation energies, the uncertainty estimates are more challenging.
Because the excitation energies in medium-mass nuclei are small compared to the
total energy scale, and because the LO interaction performs poorly in most nuclei
(as expected with only two LECs), the theoretical uncertainty would be dominated by the
large difference $\abs{E_{\text{LO}}^{\text{ex}}-E_{\text{NLO}}^{\text{ex}}}$, if we
were to follow the same prescription for the excited states as for the ground-state
energies above. We therefore adopt the following to estimate the uncertainties for
the excitation energies
\begin{align}
\Delta E_{\text{LO}}^{\text{ex}} &= \max\bigl(E^{\text{av}}_{\text{sd}},\,
\abs{E_{\text{LO}}^{\text{ex}}} \bigr) \, Q^2 \,, \label{eq:errlo} \\[1mm]
\Delta E_{\text{NLO}}^{\text{ex}} &= \max\bigl(E^{\text{av}}_{\text{sd}},
\,\abs{E_{\text{NLO}}^{\text{ex}}} \bigr) \, Q^3 \,, \label{eq:errnlo}
\end{align}
where we have introduced the scale $E^{\text{av}}_{\text{sd}} = 3$~MeV, which is
taken to be approximately the average of the splittings between the $sd$-shell orbitals.
This scale sets the natural scale for excitations in the $sd$ shell.

\subsection{Ground-state energies}

\begin{figure*}[p]
\centering
\includegraphics[width=\textwidth]{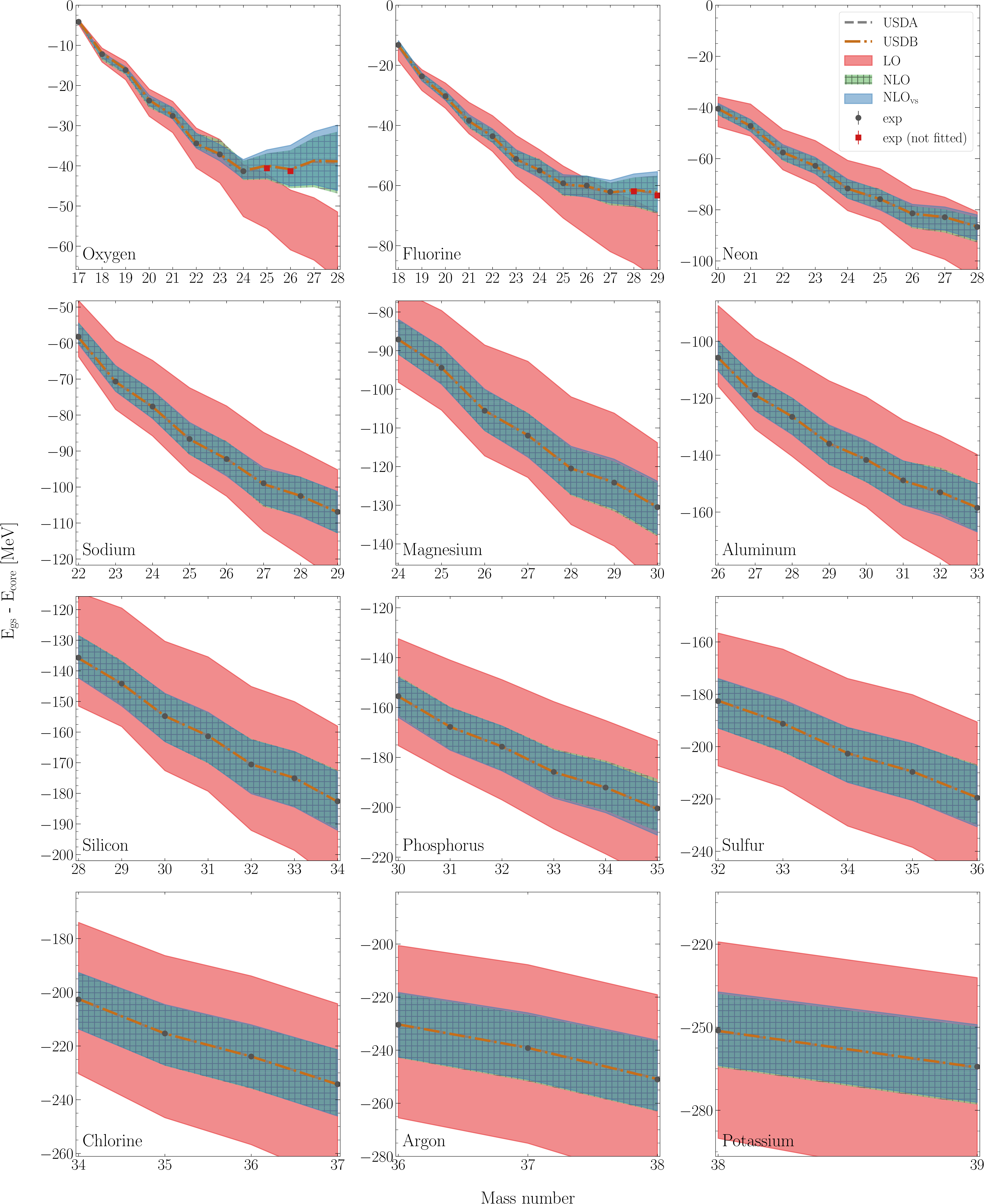}
\caption{Ground-state energies from oxygen to potassium obtained from chiral
shell-model interactions at LO, NLO, and NLO$_\mathrm{vs}$. The energies are
given with respect to the $^{16}$O core and are Coulomb corrected according
to Tab. \ref{tab:coulcorr}. The theoretical uncertainties are calculated with Eq.~\eqref{eq:errgslo}
at LO and with Eq.~\eqref{eq:errgsnlo} at NLO. For comparison, we give the USDA
and USDB results from Ref.~\cite{Brow06USD}. Experimental energies that are included
in the fit are given by gray circles, while states that are not included are given by
red squares for ${}^{25,26}$O and ${}^{28,29}$F.}
\label{fig:nuc_chains}
\end{figure*}

In Fig.~\ref{fig:nuc_chains}, we show the ground-state energies for the isotopic chains
from oxygen to potassium based on the chiral shell-model interactions at LO, NLO, and 
NLO$_\mathrm{vs}$ including the theoretical uncertainties as discussed above.
For comparison, also the USDA and USDB energies are given.
We find that all states that were included in the fit are reproduced at all orders within
the EFT uncertainties. However, the LO interaction predicts too much binding for the
neutron-rich oxygen and fluorine isotopes that were not included in the fit. As a result,
the oxygen dripline is not reproduced, being at or beyond $^{28}$O at LO, and also
$^{28}$F and $^{29}$F are overbound with respect to experiment. Remarkably,
already the NLO interaction correctly reproduces the oxygen dripline as well as the
two fluorine isotopes, which were not included in the fit. Moreover, the NLO and 
NLO$_\mathrm{vs}$ interactions overlap in all cases and reproduce ground-state
energies equally well.

\subsection{Spectra}

\begin{figure*}[p]
\centering
\includegraphics[width=.925\textwidth]{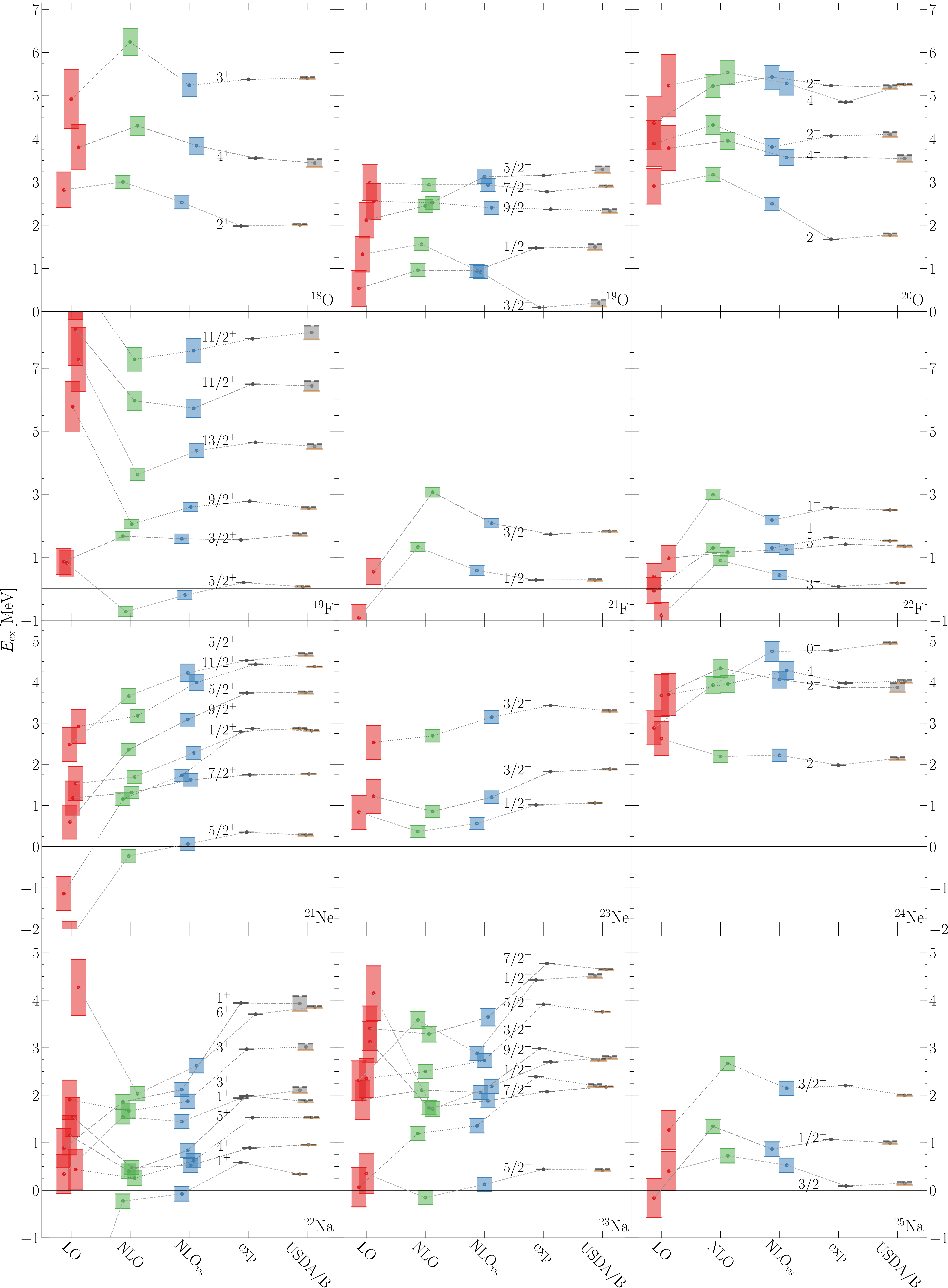}
\vspace{-.2cm}
\caption{Excitation spectra of selected isotopes from oxygen to sodium. In each
panel, results are shown for the chiral shell-model interactions at LO, NLO, and
NLO$_\mathrm{vs}$, in comparison to experiment and USDA/USDB results, where
USDA (USDB) is visualized by a dashed gray (solid orange) line. The theoretical
uncertainties are given by Eq.~\eqref{eq:errlo} at LO and by Eq.~\eqref{eq:errnlo} at
NLO. The same states are joined by dashed lines to guide the eye. The angular
momentum labels are printed next to the experimental states. Isotope labels are
given in the lower right corner.}
\label{fig:excited1}
\end{figure*}

\begin{figure*}[p]
\centering
\includegraphics[width=.925\textwidth]{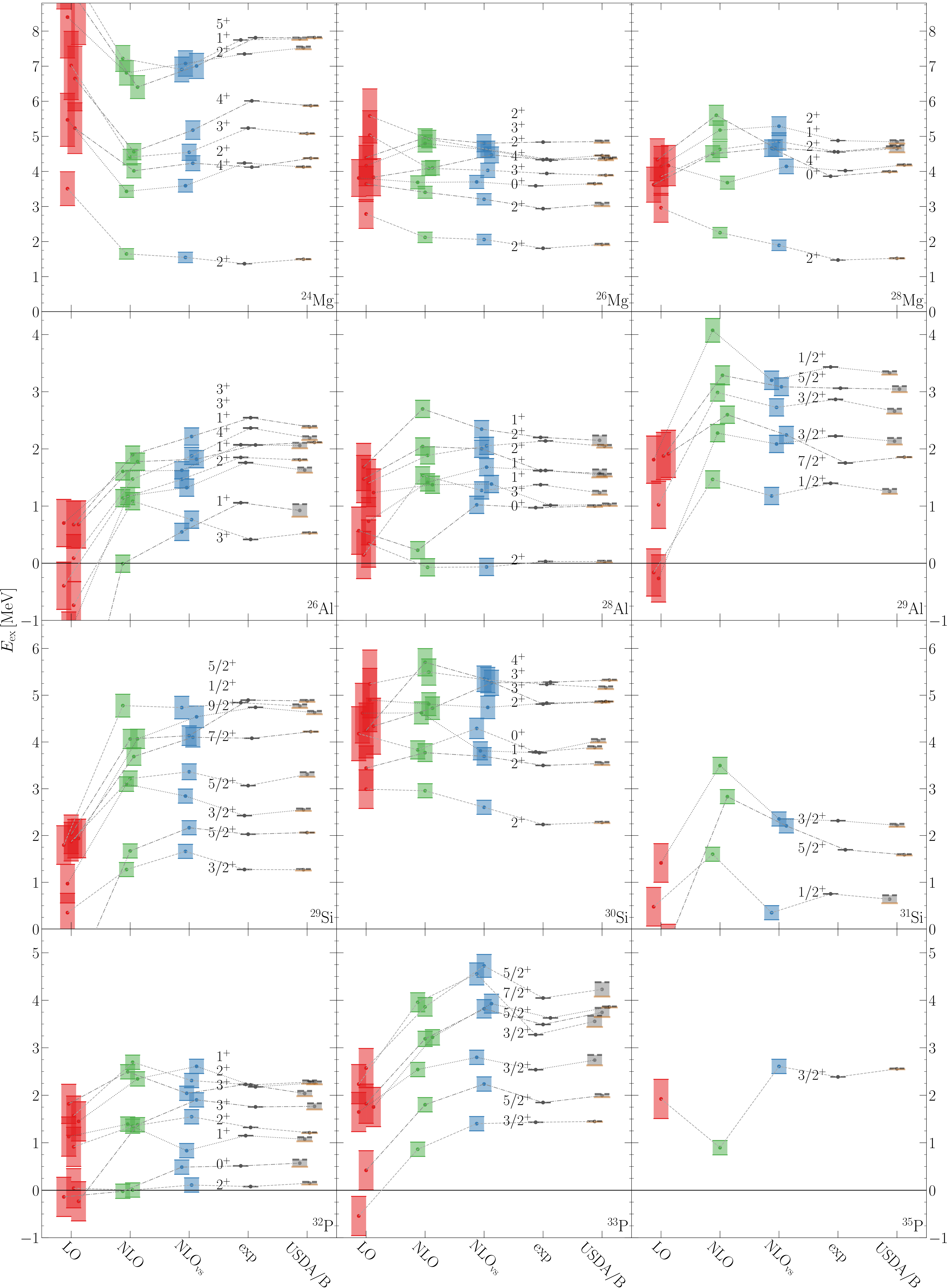}
\caption{Excitation spectra of selected isotopes from magnesium to phosphorus.
For details see the caption of Fig.~\ref{fig:excited1}.}
\label{fig:excited2}
\end{figure*}

\begin{figure*}[ht]
\centering
\includegraphics[width=\textwidth]{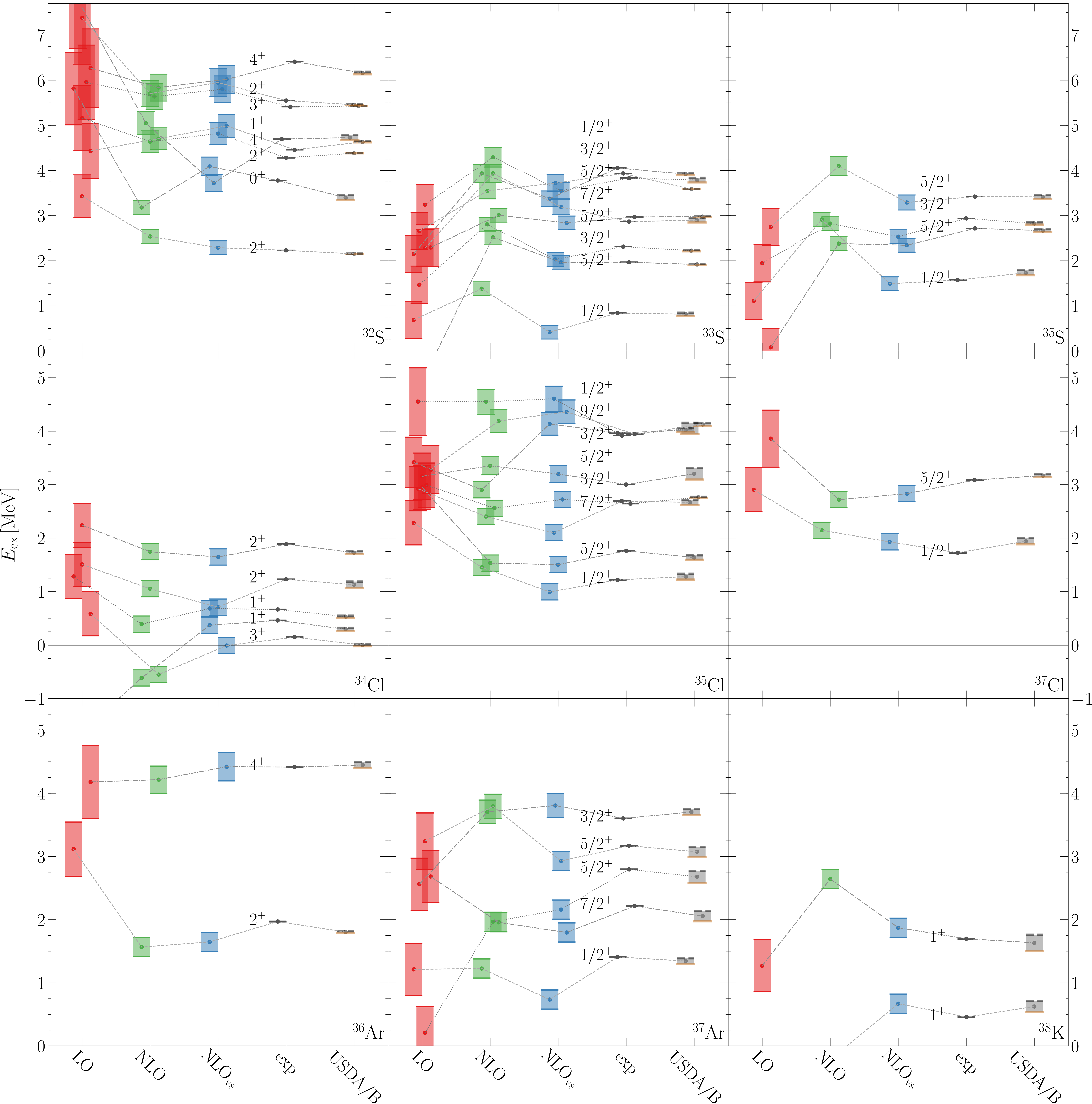}
\caption{Excitation spectra of selected isotopes from sulfur to potassium.
For details see the caption of Fig.~\ref{fig:excited1}.}
\label{fig:excited3}
\end{figure*}

In Figs.~\ref{fig:excited1}--\ref{fig:excited3}, we present our results for the spectra of excited
states. These cover the $sd$ shell for representative cases of nuclei regarding the
fits. In each panel, results are given for the chiral shell-model interactions at LO, NLO, and
NLO$_\mathrm{vs}$ including the theoretical uncertainties, in comparison to experiment
and the USDA/USDB interactions. First, in Fig.~\ref{fig:excited1}, we show results for 
oxygen, fluorine, neon, and sodium isotopes. For the oxygen spectra (first row), the excitation
energies generally change weakly from LO to NLO to NLO$_\mathrm{vs}$. From LO to NLO,
the excitation energy usually increases, and the NLO$_\mathrm{vs}$ results
generally lead to an improvement. In fluorine (second row), the NLO interaction already shows
a clear improvement from the LO result, but overshoots the experimental value somewhat,
where again then at NLO$_\mathrm{vs}$ the spectra are in good agreement with
experiment. For the neon spectra (third row), most states show a continuous
improvement from LO to NLO to NLO$_\mathrm{vs}$.
Moreover, by including the CM-dependent operators, the correct ground-state can
be reproduced in $^{21}$Ne. For the sodium isotopes (fourth row), we show the
two outliers $^{22}$Na and $^{23}$Na, which we already pointed to in the discussion
of Fig.~\ref{fig:singlenucrms}. In both of them we see that our interactions at NLO and 
NLO$_\mathrm{vs}$ lead to too low energies, and also that there is nearly no 
improvement from NLO to NLO$_\mathrm{vs}$. However, $^{25}$Na shows again a
similar behavior as the fluorine isotopes.

In Fig.~\ref{fig:excited2}, we show results for magnesium, aluminum, silicon, and
phosphorus isotopes. We generally find similar order-by-order behaviors described
above, with an overall improvement going to NLO$_\mathrm{vs}$. We
note in particular the improvement for $^{26}$Al, $^{31}$Si, and $^{35}$P, as
well as the correct reproduction of the $^{32}$P ground state when the CM-dependent
operators are included. In Fig.~\ref{fig:excited3}, we show results for the remaining
sulfur, chlorine, argon, and potassium isotopes, which exhibit similar order-by-order
trends as well. Another outlier here is the first excited state of $^{37}$Ar, which
is well reproduced by the LO and NLO interactions, but at slightly too low energy
with the NLO$_\mathrm{vs}$ interaction. However, besides this first excited state,
most of the remaining states improve with the NLO$_\mathrm{vs}$ interaction.

Finally, we need to comment on the theoretical uncertainties for the excited states.
While the behavior of the uncertainties may not be unreasonable, the adopted prescription
for the uncertainties of excitation energies is not fully satisfactory, in particular regarding
the LO to NLO behavior which is not overlapping in many cases. Future work is needed
here, with, e.g., a Bayesian analysis~\cite{Furn15uncert} of the order-by-order
behavior of the results leading to improved estimates of the theoretical uncertainties.

\subsection{Predictions}

\begin{figure*}[p]
\begin{turn}{90}
\begin{minipage}{1.41\textwidth}
\centering
\includegraphics[width=.86\textwidth]{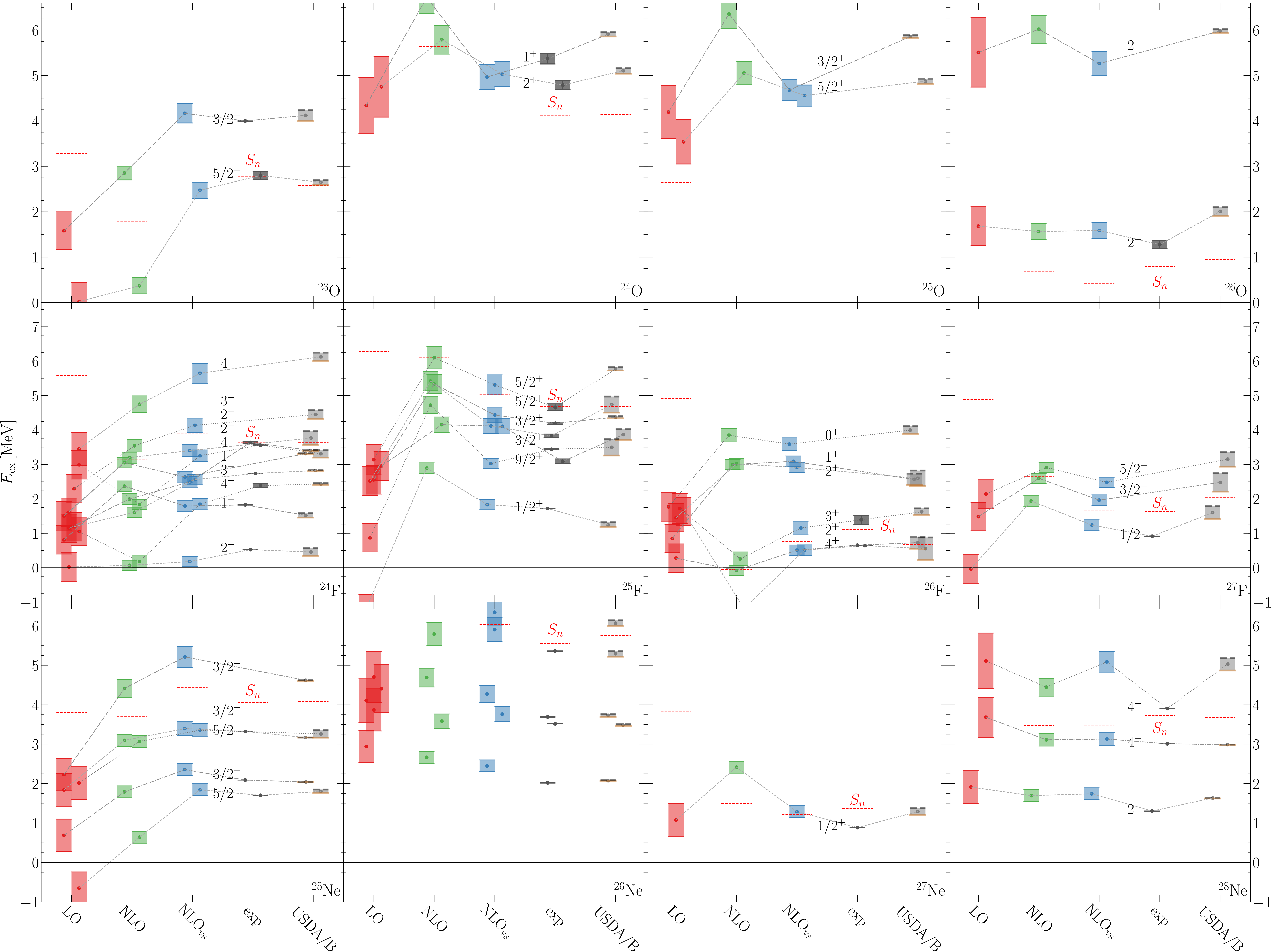}
\caption{Predictions of excited states in oxygen, fluorine, and neon isotopes.
For details on the labelling see the caption of Fig.~\ref{fig:excited1}. In addition to the
ENSDF~\cite{nndc14ENSDF} spectra, we compare our results to experimental data in
$^{23}$O~\cite{Elek0723O,Schi0723O}, $^{26}$O~\cite{Kond1626O}, 
$^{24}$F~\cite{Cace1524F}, and $^{25,26}$F~\cite{Vajt1425F,Vandebrouck172526F}.
Moreover, to emphasize the effects of the continuum, not included in our calculation,
we give the calculated and experimental neutron separation energy $S_n$ as horizontal
dashed line (see text).}
\label{fig:pred}
\end{minipage}
\end{turn}
\end{figure*}

After the promising prediction of the oxygen dripline at NLO and NLO$_\mathrm{vs}$
observed above, we also study the predictions for excited states of neutron-rich
nuclei beyond the fitted data set. We focus on the spectra of neutron-rich oxygen, 
fluorine, and neon isotopes, which are plotted in Fig.~\ref{fig:pred}. Only the first
excited state in $^{26}$Ne was included in the fit. All remaining states are predictions
of the chiral shell-model interactions. Because our calculations do not include the
continuum, we emphasize this by showing the neutron separation energy $S_n$
in Fig.~\ref{fig:pred}. For states close to or above $S_n$, the explicit inclusion
of the continuum will lead to changes, which are often of the order of few
hundred keV unless this is further resonantly enhanced.

In comparison to measured states, the chiral shell-model interactions at NLO$_\mathrm{vs}$
again lead to the best overall agreement, and there is generally an improvement
in going from LO to NLO to NLO$_\mathrm{vs}$. For the oxygen isotopes this is
especially visible in $^{23,24}$O. Moreover, all our interactions reproduce the first
2$^+$ energy in $^{26}$O recently measured at RIKEN~\cite{Kond1626O}.
This state is especially impressive, since neither the ground-state energy, nor the
excitation energy was used in our dataset, and the order-by-order behavior is very
stable. The agreement of our chiral shell-model interaction predictions at NLO$_\mathrm{vs}$
is also very good for the fluorine isotopes, especially for the low-lying states known,
and for all neon isotopes shown.

\section{Summary and outlook}
\label{sec:sum}

We have developed chiral shell-model interactions in the $sd$ shell, by fitting the LECs
of chiral EFT operators at LO and NLO directly to 441 ground- and excited-state energies.
In addition to the free-space contact interactions and pion exchanges,
this includes novel CM-dependent operators that arise due to the breaking of Galilean
invariance in the presence of the core.

The shell-model fits lead to a systematic improvement from
LO to NLO and NLO$_\mathrm{vs}$ and resulted in natural LECs at all orders. The RMS
derivation of the fits improved from 1.8~MeV at LO to 0.7~MeV at NLO and 0.5~MeV
at NLO$_\mathrm{vs}$. The latter includes five novel operators that depend on the two-body CM
momentum, so that the total number of LECs at NLO$_\mathrm{vs}$ is 14. In comparison
to USD-type interactions, the RMS deviation is about 200~keV higher, but shows a similar
rapid improvement with the number of LECs. Moreover, the monopole matrix elements are
similar to the successful USDA/USDB interactions at NLO and NLO$_\mathrm{vs}$.
Therefore, we conclude that the chiral EFT operators efficiently capture the relevant physics
at low energies. The EFT expansion enabled us to provide theoretical uncertainties, 
which seem very systematic for ground-state energies, but require further developments
for the excitation energies.

We have found a very good reproduction of experimental ground-state energies at all
orders, and a striking improvement in the reproduction of excitation energies from LO to
NLO and NLO$_\mathrm{vs}$. Moreover, the overall systematic improvement from NLO
to NLO$_\mathrm{vs}$ and the fact that some states could only be reproduced, e.g., with
the correct ground-state spin at NLO$_\mathrm{vs}$, confirms the importance of the
inclusion of the new CM-dependent operators for chiral shell-model interactions.
The developed interactions show promising predictions for
neutron-rich isotopes beyond the fitted data set. In addition to the correct reproduction
of the oxygen dripline already at NLO, the excited states in neutron-rich oxygen, fluorine,
and neon isotopes, which were not included in the fit, are predicted very well
at NLO$_\mathrm{vs}$.

Besides improving the way theoretical uncertainties can be assessed for
excited states, future work includes going to higher order, which will include also
three-nucleon interactions, and the exploration of consistent electroweak transition
based on chiral EFT operators. Moreover, for valence-space Hamiltonians beyond
a single major shell, where phenomenological interactions involve ad hoc reductions
of the cross-shell matrix elements, the strategy presented here could provide
interesting new interactions. This is especially important for exotic nuclei and
for heavier nuclei, including key nuclei for neutrinoless
double-beta decay.

\section*{Acknowledgments}

We thank G.\ F.\ Bertsch, H.\ Feldmeier, K.\ Hebeler, S.\ K\"onig, and I.\ Tews for useful discussions
and B.\ A.\ Brown for communicating to us the USDA/B experimental data set. This work was
supported by the ERC Grant No. 307986 STRONGINT, the GSI-TU Darmstadt cooperation,
and the BMBF under Contract No. 05P15RDFN1.

\appendix

\section{Partial-wave decomposition}
\label{sec:apppwd}

\subsection{Free-space contact interactions}
\label{sec:freepwd}

In the following, we give the partial-wave decomposition of the LO and NLO contact interactions, where
\begin{equation}
\braket{\vec{p}|\o{V}_{C_i}|\vec{p}'} \longrightarrow C_i \, V_{C_i}^{ll'sj}(p,p')\,,
\end{equation}
when projected on a given partial wave. The free-space contact interactions lead to
\begin{align}
&V_{C_S}^{ll'sj}(p,p') = 4\pi\del{ll'}\del{l0}\,,\\
&V_{C_T}^{ll'sj}(p,p') = 4\pi\del{ll'}\del{l0} \bigl(2s(s+1)-3\bigr)\,,\\
&V_{C_1}^{ll'sj}(p,p') = 4\pi\del{ll'} \Bigl(\del{l0}(p^2+p'^2)-\frac{2}{3}\del{l1}pp'\Bigr)\,,\\
&V_{C_2}^{ll'sj}(p,p') = \pi\del{ll'} \Bigl(\del{l0}(p^2+p'^2)+\frac{2}{3}\del{l1}pp'\Bigr)\,,\\
&V_{C_3}^{ll'sj}(p,p') = \bigl(2s(s+1)-3\bigr)V_{C_1}^{ll'sj}(p,p')\,,\\
&V_{C_4}^{ll'sj}(p,p') = \bigl(2s(s+1)-3\bigr)V_{C_2}^{ll'sj}(p,p')\,,\\
&V_{C_5}^{ll'sj}(p,p') = \del{ll'}\del{s1}\del{l1}\frac{2\pi}{3}\bigl(4-j(j+1)\bigr)pp'\,,\\
&V_{C_6}^{ll'sj}(p,p') =\nonumber\\
& 24\pi  \sum\limits_{a=0,2}(-1)^{s+j+l'+a} \, \jsc{a} \, \jsc{s}^{\,2} \sj{l}{s}{j}{s}{l'}{a}\nj{1}{1}{a}{1/2}{1/2}{s}{1/2}{1/2}{s}\nonumber\\
& \times \left[\cg{a0}{1010}\left(p^2\del{la}\del{l'0} + p'^2\del{l0}\del{l'a}\right)+\frac{2}{3}\jsc{a}\,pp'\del{ll'}\del{l1}\right]\,,\\[1mm]
&V_{C_7}^{ll'sj}(p,p') =\nonumber\\
& 6\pi  \sum\limits_{a=0,2}(-1)^{s+j+l'+a} \, \jsc{a} \, \jsc{s}^{\,2} \sj{l}{s}{j}{s}{l'}{a}\nj{1}{1}{a}{1/2}{1/2}{s}{1/2}{1/2}{s}\nonumber\\
&\times\left[\cg{a0}{1010}\left(p^2\del{la}\del{l'0}+ p'^2\del{l0}\del{l'a}\right) -\frac{2}{3}\jsc{a} \, pp'\del{ll'}\del{l1}\right]\,.
\end{align}
where $\jsc{a} = \sqrt{2a +1}$, $\cg{lm}{l_1m_1l_2m_2}$ is a Clebsch-Gordan coefficient,
and $\left\lbrace \cdots \right\rbrace$ are Wigner 6j- and 9j-symbols.

\subsection{Center-of-mass--dependent operators}
\label{sec:cmpwd}

Similar to the free-space contact interactions, we give the partial-wave decomposition of the
CM-dependent operators with
\begin{align*}
\braket{\vec{p},\vec{P}|\o{V}_{P_i}|\vec{p}',\vec{P}'} \longrightarrow \frac{\delta(P-P')}{PP'}P_i
\braket{\text{rel,cm}| V^J_{P_i} |\text{rel$'$,cm$'$}} \,,
\end{align*}
where we use the short-hand notation
\begin{equation}
\braket{\text{rel,cm}| V^J |\text{rel$'$,cm$'$}}
= \braket{pP[(ls)jL]J | V | p'P[(l's')j'L']J} \,,
\end{equation}
which is diagonal in $P$, since the total momentum is conserved, and diagonal in total
angular momentum ${\bf J} = {\bf j} + {\bf L}$, but not diagonal in $s,s'$ or $j,j'$.
With this, the partial-wave decomposition reads
\begin{align}
&\braket{\text{rel,cm}| V^J_{P_1} |\text{rel$'$,cm$'$}}  = 4 \pi P^2 \del{ss'} \del{ll'}\del{l0} \del{jj'}\del{js}\del{LL'}\,,\\
&\braket{\text{rel,cm}| V^J_{P_2} |\text{rel$'$,cm$'$}}  =  4 \pi P^2 \del{ss'} \del{ll'}\del{l0} \del{jj'}\del{js}\del{LL'}\nonumber\\
&\quad \times \bigl(2s(s+1)-3\bigr)\,,\\[1mm]
&\braket{\text{rel,cm}| V^J_{P_3} |\text{rel$'$,cm$'$}} =  24\pi\sqrt{2} \, \del{s+s',1} \, \jsc{j}\jsc{j'}\jsc{L'}\nonumber\\
&\quad \times \cg{L0}{L'010}\sj{j}{L}{J}{L'}{j'}{1}\nj{1}{1}{1}{l}{s}{j}{l'}{s'}{j'} \nonumber\\ 
&\quad \times  (-1)^{L+J+j'+s'+1}\left(\del{l1}\del{l'0}p+\del{l0}\del{l'1}p'\right)P\,,\\[1mm]
&\braket{\text{rel,cm}| V^J_{P_4} |\text{rel$'$,cm$'$}} = 12\pi\sqrt{2} \, \del{s+s',1} \, \jsc{j}\jsc{j'}\jsc{L'}\nonumber\\
&\quad \times \cg{L0}{L'010}\sj{j}{L}{J}{L'}{j'}{1}\nj{1}{1}{1}{l}{s}{j}{l'}{s'}{j'} \nonumber\\ 
&\quad \times (-1)^{L+J+j'}\left(\del{l1}\del{l'0}p-\del{l0}\del{l'1}p'\right)P\,,
\intertext{and}
&\braket{\text{rel,cm}| V^J_{P_5} |\text{rel$'$,cm$'$}} = 24\pi\del{ss'}\del{ll'}\del{l0}\del{jj'}\jsc{L'} \, \jsc{s}^{\,2} \nonumber\\
&\quad \times\sum\limits_{a=0,2}\jsc{a} \, \cg{a0}{1010} \cg{L0}{L'0a0} \sj{j}{L}{J}{L'}{j}{a}\nj{1}{1}{a}{1/2}{1/2}{s}{1/2}{1/2}{s} \nonumber\\ 
&\quad \times  (-1)^{L+J+j}P^2 \,.
\end{align}

\section{Transformation to HO basis}
\label{sec:detailsho}

To calculate the TBMEs, we first transform the partial-wave momentum-space matrix elements
to the relative-CM HO basis
\begin{align}
&\braket{nN [(ls)j L]{J} |V| n'N'[(l's')j'L']{J}} \nonumber\\
&=\int dp p^2  R_{nl}(p) \int dp' p'^2  R_{n'l'}(p') \nonumber\\
&\times \int dP P^2 R_{NL}(P) R_{N'L'}(P) \braket{\text{rel,cm}| V^J |\text{rel$'$,cm$'$}} \,.
\end{align}
The relative-CM HO matrix elements are then transformed to the TBMEs using the
Talmi-Moshinsky transformation. Including also the isospin part, this leads to
\begin{align}
&\braket{nN[(ls)jL]J T | V |n'N'[(l's')j'L']J T} \rightarrow \nonumber\\[2mm]
&\braket{(n_1 l_1 j_1)(n_2 l_2 j_2) J T | V | (n'_1 l'_1 j'_1)(n'_2 l'_2 j'_2) J T} = \nonumber\\[1mm]
&\Braket{n_1 n_2 \Bigl[ (l_1 \frac{1}{2}) j_1 (l_2 \frac{1}{2}) j_2 \Bigr] J T | V | n'_1 n'_2 \Bigl[ 
(l'_1 \frac{1}{2}) j'_1 (l'_2 \frac{1}{2}) j'_2 \Bigr] J T} .
\end{align}
To this end, we first recouple the two-body states to $\ket{n_1 n_2 [ (l_1 l_2) {\Lambda} 
(\frac{1}{2} \frac{1}{2}) s ] J}$, with total orbital angular momentum ${\bm \Lambda} = {\bf l}_1 + {\bf l}_2$ 
or in terms of relative and CM angular momenta ${\bm \Lambda} = {\bf L} + {\bf l}$, and then to
two-body states $\ket{N n [ (L l) \Lambda s ] J}$, which can be recoupled to the
desired relative-CM states $\ket{n N [ (l s) j L] J}$. Combining this, we have for antisymmetrized,
normalized two-body states
\begin{align}
&\Ket{n_1 n_2 \Bigl[ (l_1 \frac{1}{2}) j_1 (l_2 \frac{1}{2}) j_2 \Bigr] J T}
= \sum_{\Lambda s} \jsc{\Lambda} \jsc{s} \, \jsc{j_1 } \jsc{j_2}
\nj{l_1}{l_2}{\Lambda}{1/2}{1/2}{s}{j_1}{j_2}{J} \nonumber \\
&\times \sum_{n l N L} \braket{Nn(Ll)\Lambda | n_1 n_2 (l_1 l_2) \Lambda} _{d=1} \nonumber \\
&\times \sum_j (-1)^{l+s+j} \, \jsc{\Lambda} \jsc{j} \sj{L}{l}{\Lambda}{s}{J}{j}
{\mathcal F} \ket{nN [ (l s) j L] J T} \,,
\end{align}
where ${\mathcal F} = (1-(-1)^{l+s+T})/\sqrt{2(1+\delta_{n_1n_2}\delta_{l_1l_2}\delta_{j_1j_2})}$ takes into
account the normalization and antisymmetrization of the two-body states, and
$\braket{Nn(Ll)\Lambda | n_1 n_2 (l_1 l_2) \Lambda}_{d=1}$ is the 
Talmi-Moshinsky bracket in the conventions of Ref.~\cite{Kamu01TalmiMos}.
Note that for calculating the TMBEs the sum is over all $s,s'$, $j,j'$, $N,N'$, and 
$L,L'$, contrary to the case for free-space interactions when these are diagonal.

\section{Two-body matrix elements}
\label{sec:tbmes}

For completeness, we list the TBMEs of the NLO$_\mathrm{vs}$ interaction in Table~\ref{tab:TBMEs}.
The corresponding SPEs are $\varepsilon_{0d_{5/2}} =-4.14308\,\mathrm{MeV}$, $\varepsilon_{1s_{1/2}}
= -3.27235\,\mathrm{MeV}$, and $\varepsilon_{0d_{3/2}} =0.94172\,\mathrm{MeV}$, taken from
the spectrum of $^{17}$O.

\begin{table*}[t]
\caption{Two-body matrix elements for the NLO$_\mathrm{vs}$ interaction
for a given isospin $T$ and total angular momentum $J$. The single-particle
orbits are labelled by $2j_a \, 2j_b \, 2j_c \, 2j_d$.}
\label{tab:TBMEs}
{\renewcommand{\arraystretch}{1.3}
\begin{ruledtabular}
\begin{tabular}{ccrrrrrr}
orbitals&\multicolumn{1}{c}{$T$}&
\multicolumn{1}{c}{$J=0$}&\multicolumn{1}{c}{$J=1$}&
\multicolumn{1}{c}{$J=2$}&\multicolumn{1}{c}{$J=3$}&
\multicolumn{1}{c}{$J=4$}&\multicolumn{1}{c}{$J=5$}\\
\hline
\multirow{2}{*}{$5555$} & 0 & --& $-2.962356$ & --& $-1.348477$& --& $-4.366727$\\
					   & 1 & $-1.561786$& --& $-0.591304$& --& $-0.109698$& --\\ 
\multirow{2}{*}{$5553$}& 0 & -- & 2.203146& --&  1.185351& -- & --\\ 
				      & 1 &  -- & --& $ -0.782881$& --& $-0.958506$ & --\\ 
\multirow{2}{*}{$5551$} & 0 & --& --& --& $-1.671394$& --& --\\
					   & 1 & --& --& $-0.744657$& --& --& --\\ 
\multirow{2}{*}{$5533$} & 0 & --& 0.893444& --&  0.528466& --& --\\
					   & 1 & $-5.086935 $& --& $ -1.053896$& --& --& --\\ 
\multirow{2}{*}{$5531$} & 0 & --& 0.568667& --& --& --& --\\
					   & 1 &--& --& 0.908236& --& --& --\\ 
\multirow{2}{*}{$5511$} & 0 & --& $-0.106744$& --& --& --& --\\
					   & 1 & $-1.483464$& --& --& --& --& -- \\			
\multirow{2}{*}{$5353$} & 0 &--&$-5.783199$& $-4.103275$& $-2.055604$& $-4.881107$& --\\
					   & 1 & --&0.302405& 0.624787& $-0.239897$& $-0.866801$& -- \\		 		   
\multirow{2}{*}{$5351$} & 0 &--& --& $-1.513283$& 1.363597& --& --\\
					   & 1 & --& --& $-0.029187$& 0.039135& --& -- \\							   
\multirow{2}{*}{$5333$} & 0 &--& $-1.067963$& --& 1.696282& --& --\\
					   & 1 & --& --& $-0.728416$& --& --& -- \\						   
\multirow{2}{*}{$5331$} & 0 &--&1.558985& $-2.312917$& --& --& --\\
					   & 1 & --&$0.083957$& 0.922578& --& --&-- \\							   
\multirow{2}{*}{$5311$} & 0 &--&1.400007& --& --& --& --\\
					   & 1 & --& --& --& --& --& -- \\	
\multirow{2}{*}{$5151$} & 0 &--& --& $-1.943009$& $-3.565486$& --& --\\
					   & 1 & --& --& $-0.654884$& 0.217367& --& -- \\		
\multirow{2}{*}{$5133$} & 0 &--& --& --& 0.001177& --& --\\
					   & 1 & --& --&$-0.684786$& --& --& -- \\								   
\multirow{2}{*}{$5131$} & 0 &--& --& $-1.984955$& --& --& --\\
					   & 1 & --& --& 1.534945& --& --& -- \\			
\multirow{2}{*}{$3333$} & 0 & --& $-1.635266$& --& $-2.921386$& --& --\\
					   & 1 & $-0.014457$& --& $-0.046811$& --& --& -- \\								
\multirow{2}{*}{$3331$} & 0 &--& $-0.996922$& --& --& --& --\\
					   & 1 & --& --& 0.179512& --& --& -- \\								   
\multirow{2}{*}{$3311$} & 0 & --& $-0.309804$ & --& --& --& --\\
					   & 1 & $-1.585586$& -- & --& --& --& -- \\	
\multirow{2}{*}{$3131$} & 0 &--& $-3.190265$& $-2.531388$& --& --& --\\
					   & 1 & --& 0.054273& 0.236452& --& --& -- \\		
\multirow{2}{*}{$3111$} & 0 &--& 0.139808& --& --& --& --\\
					   & 1 & --& --& --&--& --& -- \\				
\multirow{2}{*}{$1111$} & 0 & --& $-3.262572$& --& --& --& --\\
					   & 1 &$-0.385534$& --& --& --& --& -- \\
\end{tabular}
\end{ruledtabular}}
\end{table*}

\bibliographystyle{apsrev4-1}
\bibliography{strongint}

\end{document}